\documentclass[twocolumn,english,aps,prb,amsmath,amsfonts,nofootinbib]{revtex4-2}
\PassOptionsToPackage{obeyFinal}{todonotes}
\usepackage[LGR,T1]{fontenc}
\usepackage[latin9]{inputenc}
\usepackage{units}
\usepackage{textcomp}
\usepackage{todonotes}
\usepackage{amsmath}
\usepackage{amssymb}
\usepackage{graphicx}

\makeatletter

\DeclareRobustCommand{\greektext}{%
  \fontencoding{LGR}\selectfont\def\encodingdefault{LGR}}
\DeclareRobustCommand{\textgreek}[1]{\leavevmode{\greektext #1}}
\ProvideTextCommand{\~}{LGR}[1]{\char126#1}

\usepackage{marginnote}

\usepackage{babel}

\makeatother

\usepackage{babel}
\begin{document}
\title{Anomalous hardening of spin waves in cobalt/molecular-semiconductor
heterostructures reveals strongly anisotropic spinterface magnetism}
\author{J. Strohsack}
\author{A. Shumilin}
\author{H. Zhao}
\author{G. Jecl}
\author{V. V. Kabanov}
\affiliation{Department of Complex Matter, Jozef Stefan Institute, Jamova 39, 1000
Ljubljana, Slovenia}
\author{M. Benini}
\author{R. Rakshit}
\author{V. A. Dediu}
\affiliation{ISMN-CNR, Via Piero Gobetti 101, Bologna, Italy}
\author{M. Rogers}
\author{S. Ozdemir}
\author{O. Cespedes}
\affiliation{School of Physics and Astronomy, University of Leeds, Leeds, LS2 9JT,
United Kingdom}
\author{U. Parlak}
\author{M. Cinchetti}
\affiliation{Technische Universität Dortmund, Otto-Hahn-Straße 4, Dortmund, Germany}
\author{T. Mertelj}
\email{tomaz.mertelj@ijs.si}

\affiliation{Department of Complex Matter, Jozef Stefan Institute, Jamova 39, 1000
Ljubljana, Slovenia}
\affiliation{Center of Excellence on Nanoscience and Nanotechnology Nanocenter
(CENN Nanocenter), Jamova 39, 1000 Ljubljana, Slovenia}
\date{\today}
\begin{abstract}
The interface between a ferromagnetic metal and an organic molecular
semiconductor, commonly referred to as a spinterface, is an important
component for advancing spintronic technologies. Hybridization of
the ferromagnetic-metal surface $d$ orbitals with the molecular-semiconductor
$p$ orbitals induces profound modifications not only in the interfacial
molecular layer, but also in the surface ferromagnetic-metal atomic
layer. These effects are particularly pronounced at low temperatures,
manifesting as substantial modifications in the magnetic properties
of thin-film magnetic-metal/organic heterostructures. Despite extensive
research and interest, the magnetic-ordering and magnetic-properties
of the spinterface remain poorly understood. Using ultrafast time-resolved
magneto-optical spectroscopy, to investigate the magnetic dynamics
in such heterostructures, we unveil the unique spinterface-magnetism
and its universality for a broad variety of cobalt/molecular-semiconductor
interfaces. In particular, our findings demonstrate the presence of
highly anisotropic low-temperature superparamagnetism at the cobalt/molecular-semiconductor
spinterface. This anisotropic interfacial superparamagnetism is likely
driven by strong chemical modifications in the cobalt interfacial
layer caused by the chemisorbed molecular layer. These results highlight
the pivotal role of molecular chemisorption in tuning the magnetic
properties at spinterfaces, paving the way for future spintronic applications.
\end{abstract}
\maketitle

\section{introduction}

\begin{figure*}
\includegraphics[width=0.9\textwidth]{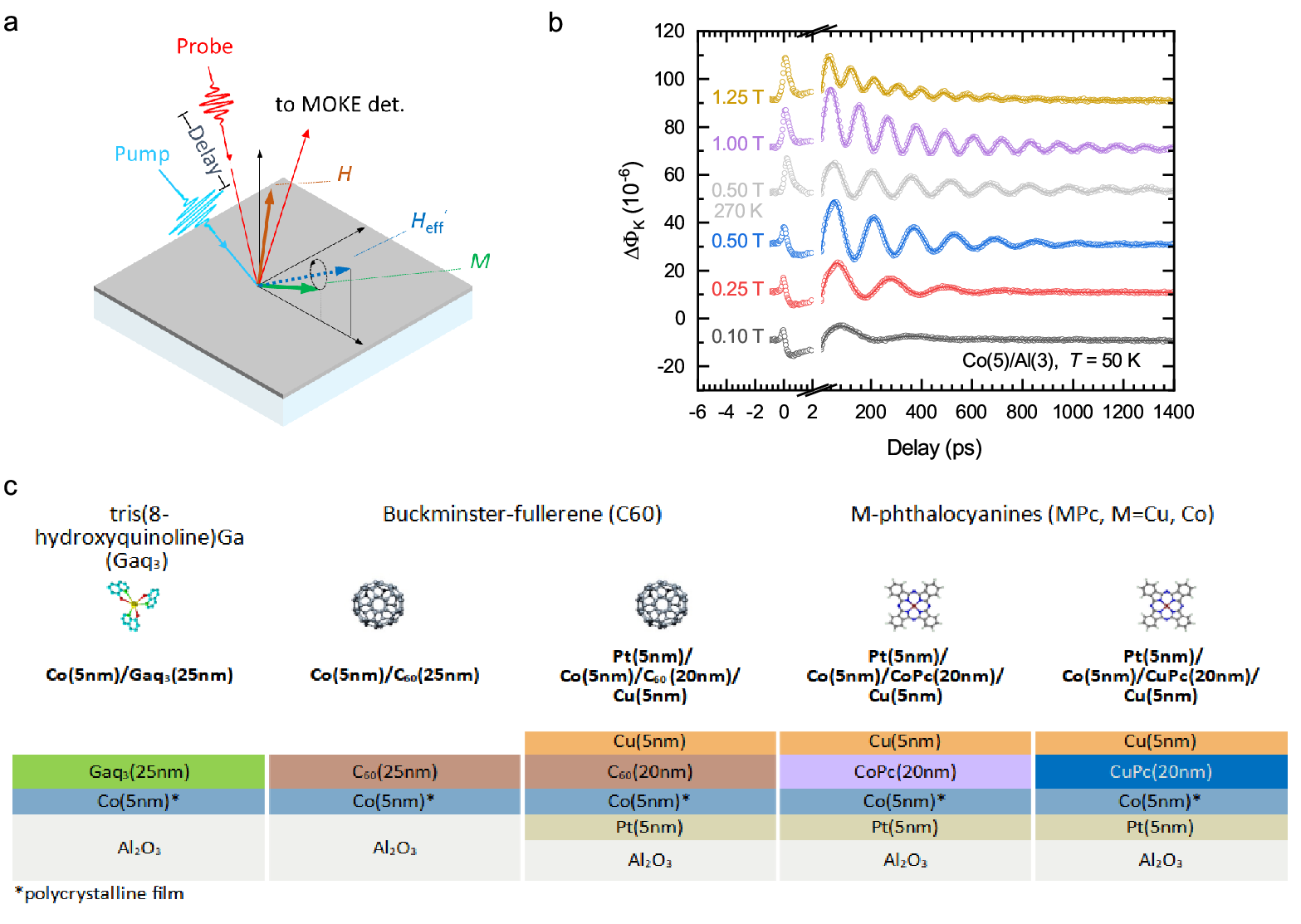}

\caption{Experimental setup and samples. \textbf{a} Schematics of the experimental
geometry. \textbf{b} Transient magneto-optical Kerr rotation in a
reference Co(5)/Al(3) sample at $T=50$~K. A trace obtained at $T=270$~K
and $\mu_{0}H=0.5$~T (gray) is shown for comparison to emphasize
weak $T$ dependence. The lines correspond to the damped oscillator
fit \eqref{eq:fitf} discussed in text. \textbf{c} Schematic representation
of the studied heterostructures.\label{fig:Experimental-setup-and}}
\end{figure*}

Hybridized interfaces between $3d$ ferromagnetic (FM) metals and
organic molecules have attracted significant interest, both for their
novel fundamental phenomena and their emergent functional properties
\citep{cinchetti2017activating}. These interfaces have been extensively
studied for more than a decade \citep{raman2013interfaceengineered,moorsom2014spinpolarized,bairagi2015tuningthe,gruber2015exchange,djeghloul2016highspin,moorsom2020pianisotropy,bairagi2018experimental,boukari2018disentangling,benini2022indepth,avedissian2022exchange,halder2023theoretical},
revealing significant alterations in the magnetic properties of thin
$3d$ ferromagnetic films upon the deposition of a single molecular
layer \citep{moorsom2014spinpolarized,bairagi2015tuningthe}. These
modifications have been attributed to the hybridization of the surface
$d$ orbitals of the FM metal with the $p$ orbitals of the deposited
molecules \citep{chen2010effectof,bairagi2015tuningthe,cinchetti2017activating,halder2023theoretical}.

Most previous investigations have been focused on the static magnetization
properties \citep{raman2013interfaceengineered,moorsom2014spinpolarized,bairagi2015tuningthe,gruber2015exchange,bairagi2018experimental,moorsom2020pianisotropy,benini2022indepth,avedissian2022exchange,halder2023theoretical,BeniniShumilin2024}
reporting increased low-temperature coercitivity. While coercitivity
is an important application parameter, its connection to the microscopic
parameters appears to be rather indirect \citep{BeniniShumilin2024}.
Moreover, the marked temperature dependence, with the effects diminishing
at room temperature, which is regularly observed \citep{moorsom2014spinpolarized,bairagi2015tuningthe,moorsom2020pianisotropy,bairagi2018experimental,benini2022indepth},
is not fully understood. In the Co/C$_{60}$ systems, for instance,
this temperature dependence has been associated \citep{moorsom2020pianisotropy}
with a particular C$_{60}$ molecule rotational transition that cannot
apply for differently shaped molecules.

To deepen our understanding, more direct and complementary data that
span a broad variety of differently shaped molecules, while accessing
the microscopic parameters such is the anisotropy and possible exchange
bias, are desirable. To provide such data one of the tools of choice
is the time-resolved magneto-optical Kerr effect spectroscopy (TR-MOKE).
It enables the detection of the magnetic dynamics on ultrafast timescales
\citep{kirilyuk2010ultrafast}. In particular, the method can be instrumental
for mapping the effective anisotropy field of thin films revealing
the microscopic magnetic free-energy parameters \citep{vankampen2002alloptical,bigot2005ultrafast}.

In this study, we present a systematic TR-MOKE investigation of several
Co/molecular-semiconductor interfaces, We compare the effects of four
molecular species with distinct molecular shapes: a highly asymmetric
tris(8-hydroxyquinoline)Ga (Gaq$_{3}$), a spherical Buckminster-fullerene
(C$_{60}$) and a couple of almost-planar M-phthalocyanine species
(MPc, M=Cu, Co). A detailed list of the thin-film heterostructures
studied is shown in Fig. \ref{fig:Experimental-setup-and} c.

We use femtosecond laser pulses to transiently perturb the equilibrium
magnetization \citep{vankampen2002alloptical,bigot2005ultrafast}
and induce coherent spin waves, which we detect with an appropriately
delayed probe pulses, as illustrated schematically in Fig. \ref{fig:Experimental-setup-and}a.
By varying the external magnetic field, $H$, in the quasi-polar MOKE
configuration, we map the effective anisotropy field, $H_{\mathrm{eff}}$,
across a range of the out-of-plane magnetization angles to deduce
the parameters of the effective magnetic free-energy terms.

The experiment reveals a distinctive hardening of the spin-wave frequency
as the temperature is lowered below $T\sim200$~K. The spin-wave
frequency external-magnetic-field dependencies indicate that the the
Co/molecular-semiconductor interface induces a suppression of the
effective Co-film easy-plane anisotropy, along with a significant
in-plane exchange-bias-like anisotropy with the corresponding effective
field on the order of $\mu_{0}H_{\mathrm{b}}\sim0.6$~T at low $T$.

Surprisingly, the behavior is found to be very similar for different
molecular species/shapes and is attributed to the presence of a strongly-anisotropic
super-paramagnetic ordering at the hybridized Co-molecules interface
(spinterface) that sets in below $170-200$~K.

\begin{figure*}
\includegraphics[width=0.9\textwidth]{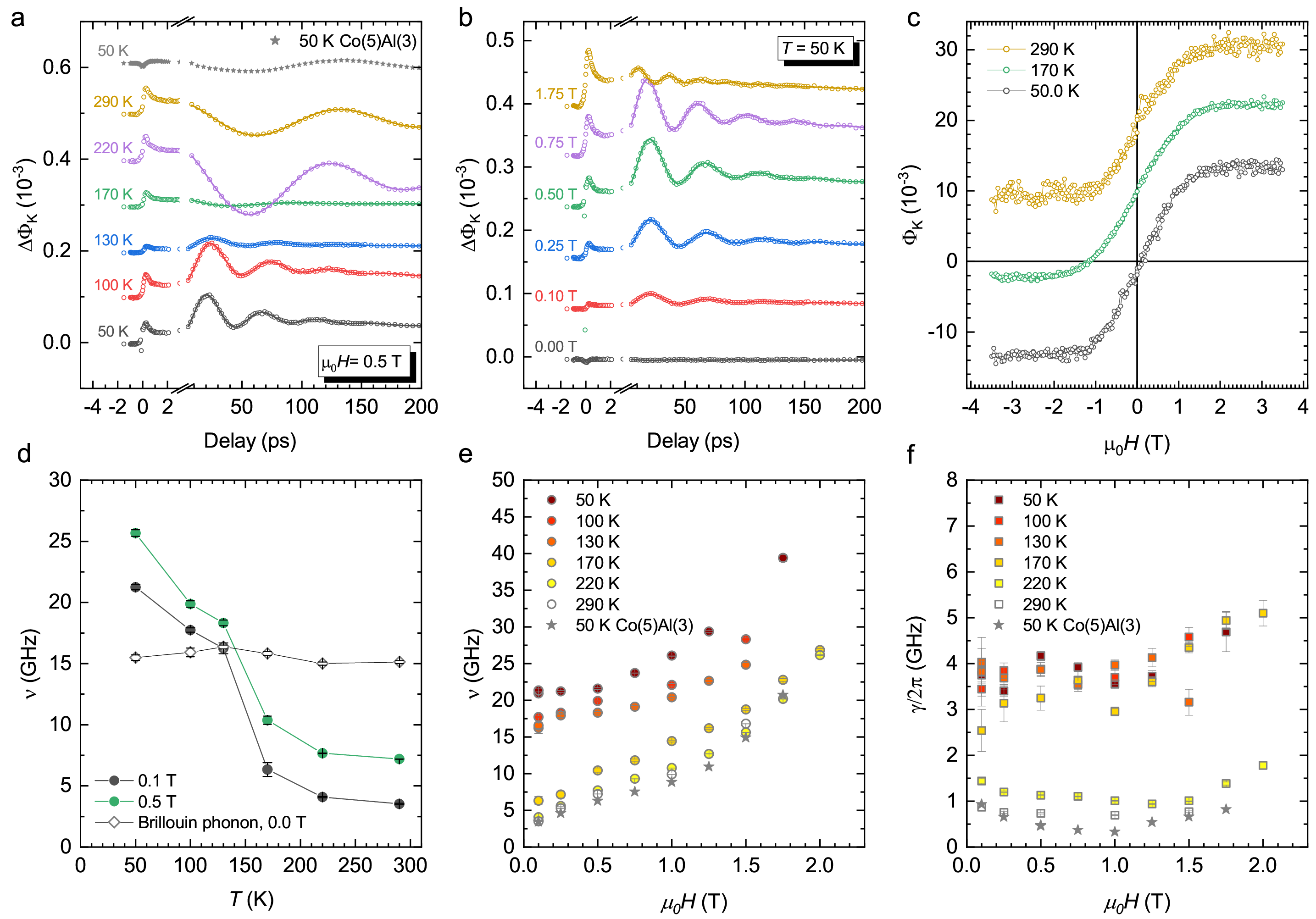}

\caption{Transient and static MOKE in Pt(5)/Co(5)/CoPc(20)/Cu(5) heterostructure.
\textbf{a} MOKE transients temperature dependence at $\mu_{0}H=0.5$
T. \textbf{b} Magnetic field dependence of MOKE transients at $T=50$
K. \textbf{c} The static Kerr rotation loops as a function of $T$.
\textbf{d} The spin wave frequency as a function of $T$. The open
symbols correspond to the coherent Brillouin waves observed in transient
reflectivity (see Supplemental, Fig. \ref{fig:Transient-reflectivity}).
\textbf{e }The spin wave frequency as a function of $H$ at different
$T$. \textbf{f} The spin wave dephasing as a function of $H$ at
different $T$. The traces in \textbf{a},\textbf{ b} and \textbf{c}
are vertically offset for clarity. The lines in \textbf{a} and \textbf{b
}are damped oscillator fits (see Supplemental, Eq. \eqref{eq:fitf}).\label{fig:Transient-and-static}}
\end{figure*}

\section{Results}

\begin{figure}
\includegraphics[width=0.9\columnwidth]{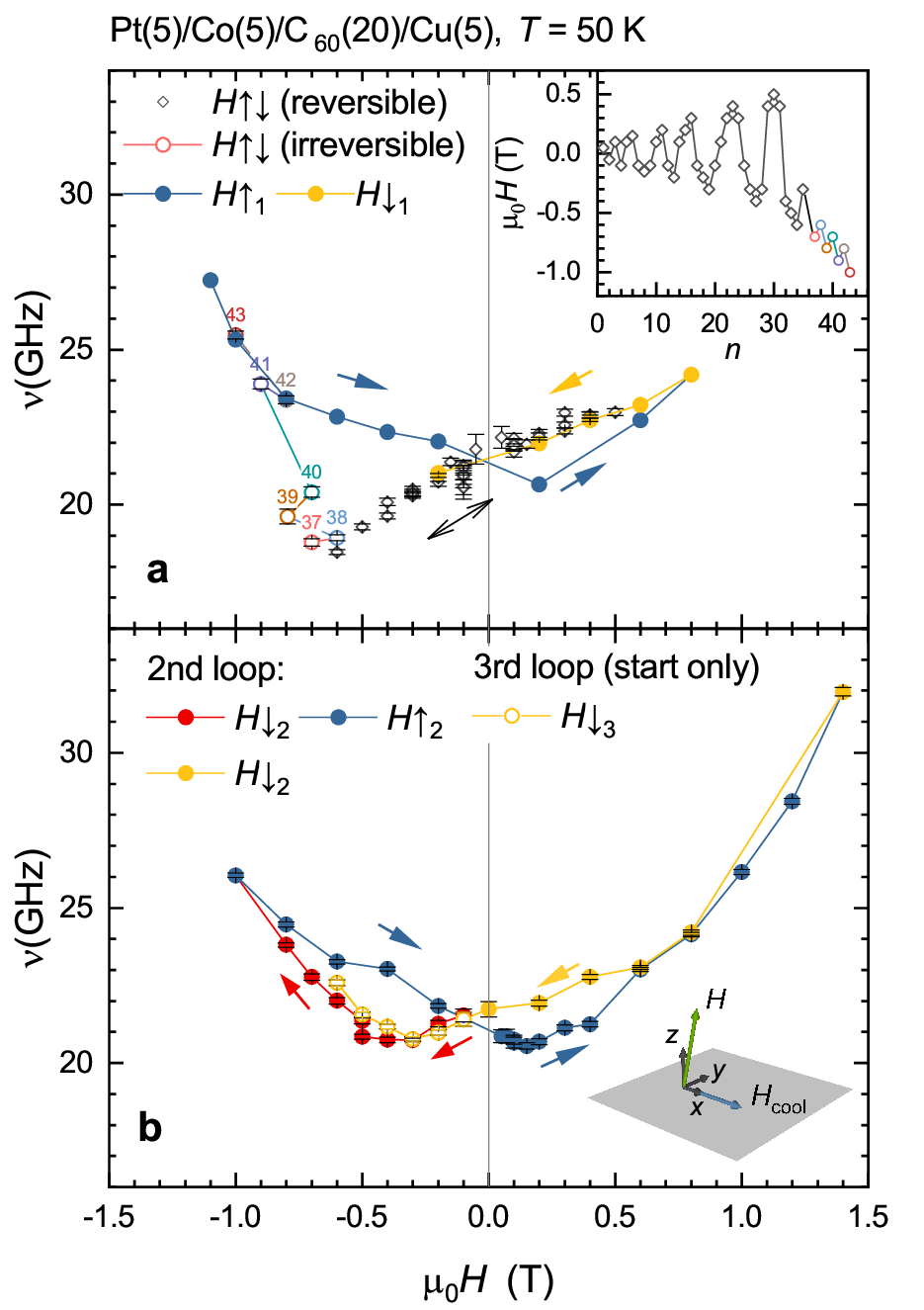}

\caption{Magnetic field dependence of the spin-wave frequency in the field-cooled
(FC) (in-plane, $\mu_{0}H_{\mathrm{cool}}=5$~T) Pt(5)/Co(5)/C$_{60}$(20)/Cu(5)
heterostructure. \textbf{a} The initial-branch reversible region (open
diamonds) and the 1st irreversible loop. The inital oscillating magnetic-field
sequence, corresponding to the open symbols, is shown in the inset.
The full symbols represent subsequent measurements where the magnetic
field was swept monotonously from -1 T to 0.8 T and back to -0.2~T.
The number labels indicate the sequential measurements to emphasize
the non-monotonous field sequence in the first part of the loop. \textbf{b}
The 2nd and the beginning of the 3rd loop measured after the 1st loop.
The magnetic field was swept in a monotonous manner: -0.1~T$\rightarrow$-1~T$\rightarrow$1.4
T$\rightarrow$-0.6 T. \label{fig:FC-hyst}}
\end{figure}

\begin{figure*}
\includegraphics[width=0.8\textwidth]{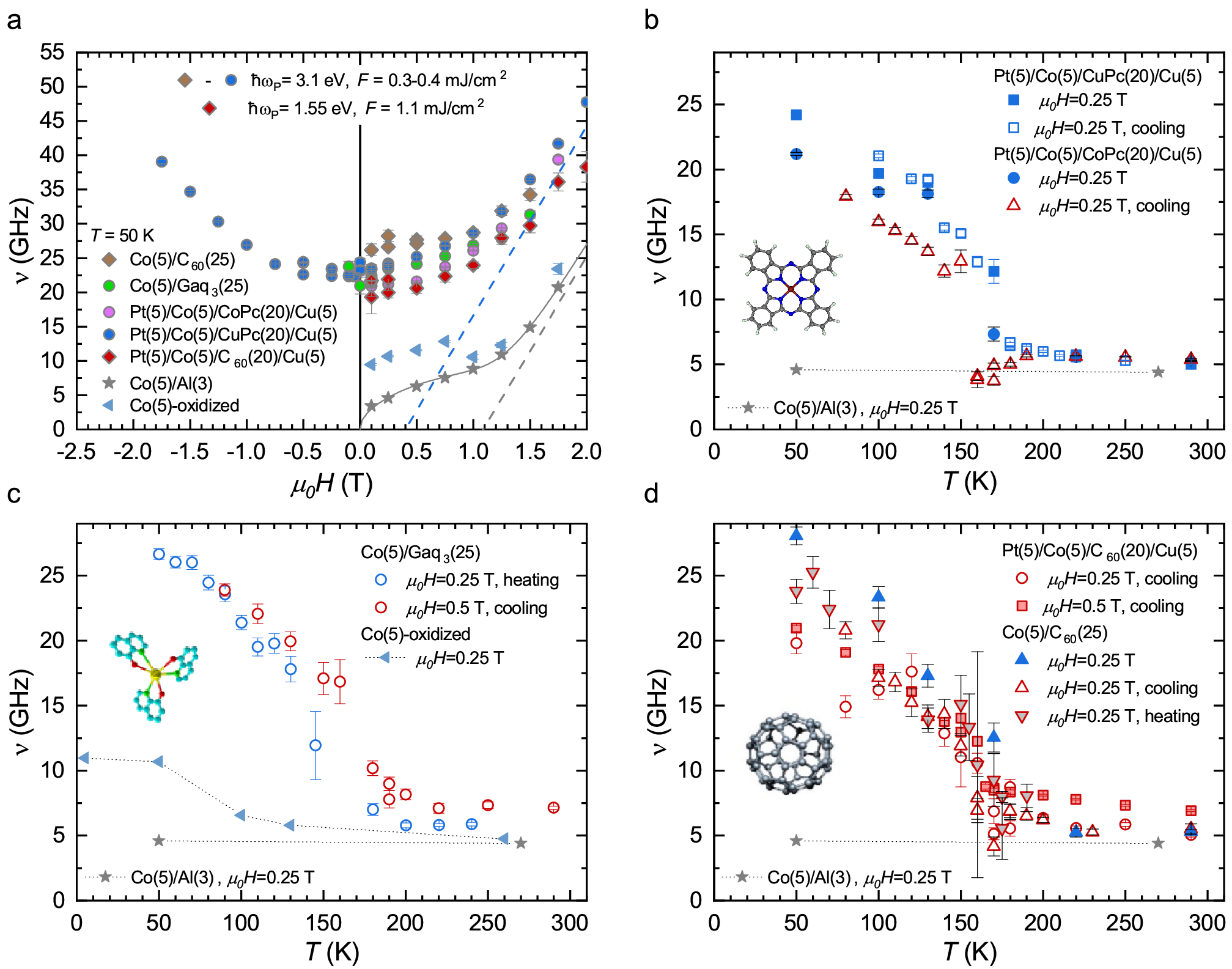}\caption{The spin-wave frequency as function of $H$ and $T$ in different
heterostructures. \textbf{a} The frequency as a function of external
magnetic field at $T=50$~K in all studied samples upon zero field
cooling. The continuous gray line is the uniform-precession easy-plane
fit (see Eq. \eqref{mod1} in Supplemental, Section \ref{subsec:Uniform-precession-modeling})
to the Co(5)/Al(3) sample data. The dashed lines indicate the large-$H$
asymptotics. \textbf{b }The frequency as a function of temperature
at a constant external magnetic field in samples with CoPc and CuPc.
\textbf{c }The frequency as a function of temperature at a constant
external magnetic field in the sample with Gaq$_{3}$. \textbf{d }The
frequency as a function of temperature at a constant external magnetic
field in samples with C$_{60}$. The full single color symbols in
\textbf{c}, \textbf{b} and \textbf{d} correspond to the data points
obtained from the zero-field-cooling isothermal $H$-scans at different
$T$, while the nonuniform symbols correspond to the constant-$H$
temperature scans. The blue symbols correspond to $\hbar\omega_{\mathrm{P}}=3.1$
eV and $F=0.3-0.4$~mJ/cm$^{2}$ while the symbols with a red border
correspond to $\hbar\omega_{\mathrm{P}}=1.55$ eV and $F=1.1$~mJ/cm$^{2}$.\label{fig:Frequency-as-function}}
\end{figure*}

In Figs. \ref{fig:Transient-and-static} a and b we show, as an example,
the temperature and magnetic field dependent transient magnetooptical
Kerr angle, $\Delta\Phi_{\mathrm{K}}$, in the Pt(5)/Co(5)/CoPc(20)/Cu(5)\footnote{Numbers in parentheses correspond to the thickness of the layers given
in nm.} heterostructure. $\Delta\Phi_{\mathrm{K}}$ is characterized by the
presence of coherent oscillations that correspond to the spin waves
\citep{vankampen2002alloptical,bigot2005ultrafast}. The comparison
to the reference Co(5)/Al(3), where Al layer is used to protect the
Co film from oxidation, shows similar spin waves at room temperature.
Upon cooling, the reference transients show only minor change (see
Fig. \ref{fig:Experimental-setup-and}b) while the Co/organic-semiconductor\textit{
}heterostructures \textit{show a remarkable hardening of the spin
wave frequency} (Fig. \ref{fig:Transient-and-static}d and e), increased
dephasing (Fig. \ref{fig:Transient-and-static}f) and a $\Delta\Phi_{\mathrm{K}}$
sign change below $T\sim170$ K (Fig. \ref{fig:Transient-and-static}a).
Other samples from Fig. \ref{fig:Experimental-setup-and} c show qualitatively
similar behavior (see Supplemental, Fig. \ref{fig:Example-of-transient}).

The spin-wave frequency is external-magnetic-field, $H$, dependent
showing hysteretic behavior (see Fig. \ref{fig:FC-hyst} and Supplemental,
Fig. \ref{fig:FCvsZFC}). The hysteresis depends on the applied external
magnetic field during cooling and exhibits training behavior. In the
Pt(5)/Co(5)/C$_{60}$(20)/Cu(5) sample upon field cooling, for example,
we initially observe a reversible monotonous magnetic-field frequency
dependence down to the negative irreversibility field, $\mu_{0}H_{\mathrm{ir-}}\sim-0.7$~T
($\nu\sim19$~GHz) (see Fig. \ref{fig:FC-hyst}a, point \#37). In
the subsequent loops (Fig. \ref{fig:FC-hyst}b), after some training,
the frequency does not drop below $\nu\sim20.5$~GHz anymore and
shows an upturn with decreasing $H$ around $\mu_{0}H\sim-0.4$~T
already. The hysteresis closing is asymmetric at $\mu_{0}H_{\mathrm{c-}}\sim-0.9$~T
on the negative and $\mu_{0}H_{\mathrm{c+}}\sim0.6$~T on the positive
side, clearly showing a field-offset behavior. 

The zero-field cooling data show qualitatively similar hysteresis
(see Supplemental, Fig. \ref{fig:FCvsZFC}), albeit with a smaller
field offset, which was not studied in detail. Most of the data (Fig.
\ref{fig:Transient-and-static}e and Fig. \ref{fig:Frequency-as-function}a)
was acquired upon zero-field cooling without reversing $H$. Such
scans showed mostly reversible behavior when sweeping $H$ from small
to the largest field and back. Some additional scatter sometimes observed
at low $H$ can be attributed to the presence of the hysteresis.

The dephasing shows only a weak magnetic-field dependence while increasing
\ensuremath{\sim}3-4 times with decreasing $T$ below $\sim170$
K (Fig. \ref{fig:Transient-and-static}f). Comparing the low-$T$
$H$-dependence for different molecules we observe qualitatively similar
behavior with the low-$H$ spin-wave frequency in the 20-30 GHz range,
which is $\sim10$ times larger than the room-$T$ (as well as the
Co(5)/Al(3) reference sample) value (Fig. \ref{fig:Transient-and-static}e
and Fig. \ref{fig:Frequency-as-function}a). The static magneto-optical
Kerr angle, corresponding to the magnetization component perpendicular
to the film, is shown in (Fig. \ref{fig:Transient-and-static}c).
Within the experimental accuracy the static Kerr angle does not show
any $T$ dependence nor a clear hysteresis.

The low-$H$ spin-wave frequency $T$-dependence is summarized in
Fig. \ref{fig:Frequency-as-function}b-d. The behavior is qualitatively
similar in all studied heterostructures. The spin-wave frequency hardening
with decreasing $T$ sets in around $T\sim200$ K with a more pronounced
jump in the 150-170 K range (depending on the molecular species) concurrently
with the oscillation sign change. The effect does not depend significantly
on the details of the auxiliary heterostructure layers as indicated
by the data from two different C$_{60}$-based heterostructures (Fig.
\ref{fig:Frequency-as-function}d).

\begin{figure}
\includegraphics[width=0.8\columnwidth]{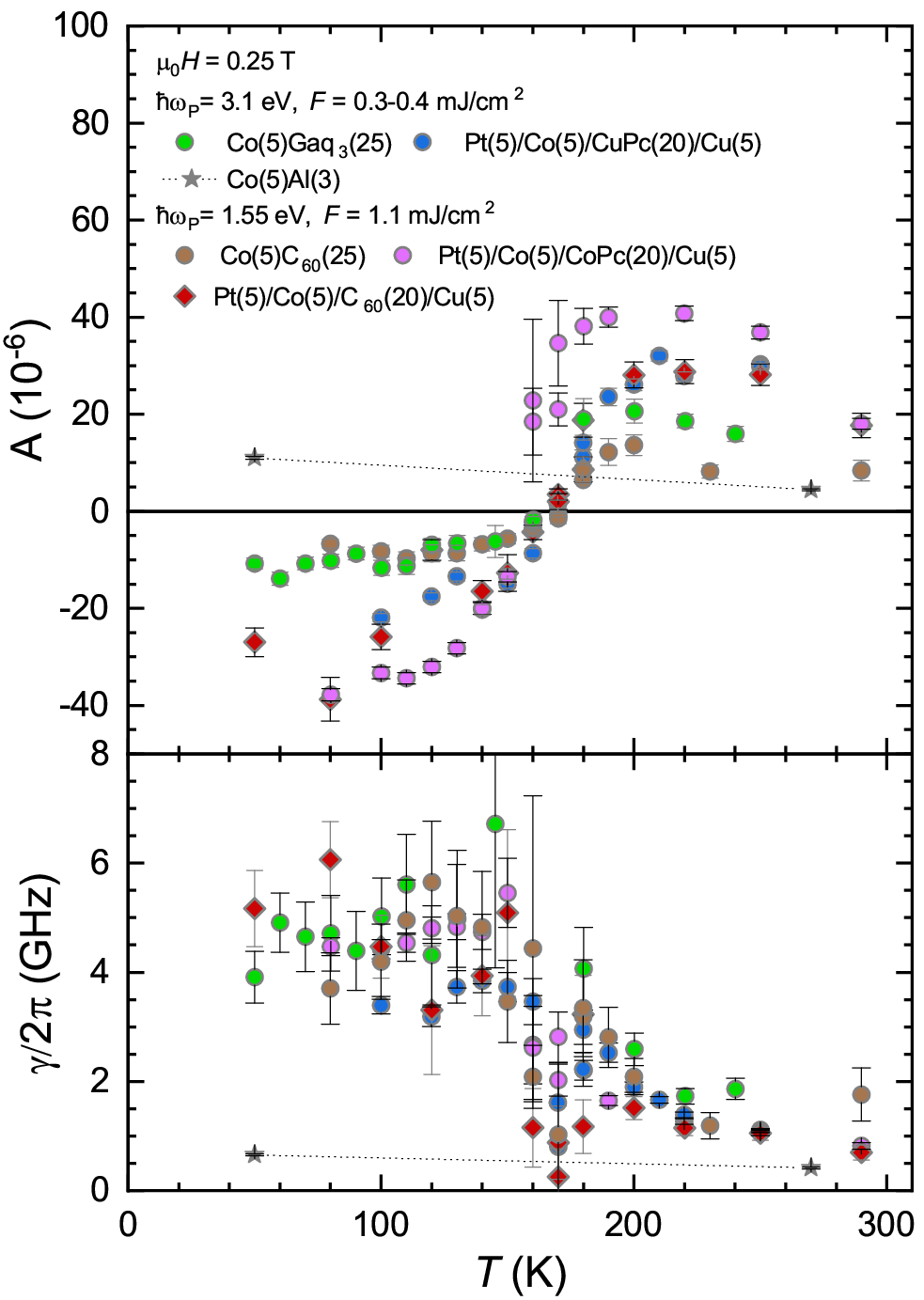}\caption{Comparison of the initial oscillation amplitude and damping as function
of $T$ in different heterostructures at $\mu_{0}H=0.25$~T.\label{fig:AvsT}}
\end{figure}

\section{Discussion}

The reference Co(5)/Al(3) heterostructure $H$-dependence of the spin-wave
frequency can be successfully modeled (the fit line in Fig.~\ref{fig:Frequency-as-function}a)
assuming a homogeneous magnetization precession with an easy plane
anisotropy\footnote{For simplicity we merge the shape and magneto-crystalline anisotropy
contributions into a single term.} of $K_{\bot}M_{0}\sim1.1$ T \textit{in the full} $T$-range (see
Supplemental, Section \ref{subsec:Uniform-precession-modeling} and
Ref. \citep{chappert1986ferromagnetic}). At high-$T$, the Co/molecules
heterostructures spin-wave frequency $H$-dependencies show similar
behaviors as the Co(5)/Al(3) reference sample (see Fig. \ref{fig:Transient-and-static}e).
The high-$T$ dynamic behavior, together with the static data (see
Fig. \ref{fig:Transient-and-static}c), is therefore consistent with
the easy plane anisotropy of $K_{\bot}M_{0}\sim1.1$ T, as in the
Co(5)/Al(3) reference sample.

At low $T,$ however, the spin-wave frequency $H$-dependencies (Fig.~\ref{fig:Frequency-as-function}a)
appear \textit{clearly incompatible} with the simple easy-plane anisotropy
model. Concurrently, the static $\Phi_{\mathrm{K}}$ data (Fig. \ref{fig:Transient-and-static}c)
show a virtually $T$-independent behavior, suggesting the easy-plane-anisotropy
model also at low-$T$. The apparent dichotomy between the static
and dynamic behavior could indicate that the observed coherent oscillations
do not correspond to the uniform spin precession at low $T$, but
to a localized mode at the Co/molecules interface or a higher order
\citep{kittel1958excitation} standing spin-wave mode. Such scenario
is, however, very unlikely. The uniform mode (with a lower frequency)
should always contribute to the response, but in the data only one
frequency component is observed in all samples (see Supplemental,
Section \ref{sec:Fit-function}, Fig. \ref{fig:Example-of-transient}).
Moreover, the optical penetration depth in Co at the probe wavelength
of $\sim13$ nm\citep{johnson1974optical} is more than twice as large
than the Co thickness. Any higher order standing modes should have
weak intensities in comparison to the uniform mode.

The uniform mode bulk excitation is a consequence of a rapid change
of the Co-film electronic and lattice temperatures \citep{bigot2005ultrafast}
and cannot be directly suppressed by the interface interaction. The
uniform mode oscillation can therefore be suppressed only if the bulk
and interface excitation contributions exactly cancel. This appears
to happen around $T\sim170$ K, where the hardening sets in and the
signal vanishes (see Fig. \ref{fig:AvsT}). It is, however, extremely
unlikely that both contributions would remain exactly balanced in
the broad range of $T$ below $T\sim170$ K in all studied samples.

It is also unlikely that the interfacial magnetization would be pinned
\citep{kittel1958excitation}, due to the high exchange coupling\footnote{The mean-field Weiss molecular field is estimated to be $\mu_{0}H_{\mathrm{W}}\sim1000$
T in Co. \citep{stohr2006magnetism}} and the small thickness of the Co film. The fundamental standing
spin-wave mode is therefore expected to be negligibly nonuniform.
As a result, the observed spin-wave oscillations correspond to the
bulk Co film dynamics, with a high degree of certainty. We can therefore
discuss the magnetization precession as being uniform along the thickness
of the Co film, where the interface induced anisotropy and/or exchange
bias can be treated as a (properly re-normalized) volume contribution.

To understand the static behavior in an in-plane external magnetic
field it was previously assumed \citep{BeniniShumilin2024} that the
effect of molecules at the interface is an induced effective in-plane
magnetic-anisotropy free-energy term, $K_{||}({\bf M}\cdot{\bf e}_{\mathrm{R}})^{2}/2$,
where $K_{||}\ll K_{\perp}$ and $K_{||}M_{0}\sim0.1$\,T, with a
random but spatially correlated in-plane vector ${\bf e}_{\mathrm{R}}$.
Although such an anisotropy can lead to a quite unusual ferromagnetic
glass state for small external magnetic fields and explain the static
measurements in an in-plane magnetic field \citep{BeniniShumilin2024},
it cannot describe the observed spin-wave frequency magnitudes. At
small external fields, where the magnetization lies in the hetero-structure
plane, the spin-wave frequency of $\sim25$ GHz (Fig. \ref{fig:Frequency-as-function}a)
corresponds to an in-plane effective field, $\mu_{0}H_{\mathrm{||eff}}\sim0.9$
T. This value is much larger than expected from the static in-plane
model parameters \citep{BeniniShumilin2024}, where \citep{sovskii2016ferromagnetic}
$\mu_{0}H_{\mathrm{||eff}}=M_{\mathrm{0}}\sqrt{K_{||}(K_{||}+K_{\perp})}\sim0.35$\,T
could be estimated from $K_{||}M_{0}\sim0.1$\,T and $K_{\perp}M_{0}\sim1.1$\,T.

The increase of the spin-wave frequency in weak external fields could
also be a consequence of the laterally inhomogeneous ferromagnetic
glass state \citep{BeniniShumilin2024}, however, the low-$T$ increase
(with respect to the Co(5)/Al(3) reference) persists in the uniformly
magnetized saturated magnetic state at large external magnetic fields.
At $\mu_{0}H=2$ T, for example, the precession frequency of $\sim40$
GHz corresponds to an out-of plane effective field of $\mu_{0}H_{\mathrm{\bot eff}}\sim1.5$
T, while a significantly smaller $\mu_{0}H_{\mathrm{\bot eff}}\sim1$
T is inferred (dominated by the easy-plane anisotropy) for the case
of the Co(5)/Al(3) reference sample at the same external field.

\begin{figure}
\includegraphics[width=0.9\columnwidth]{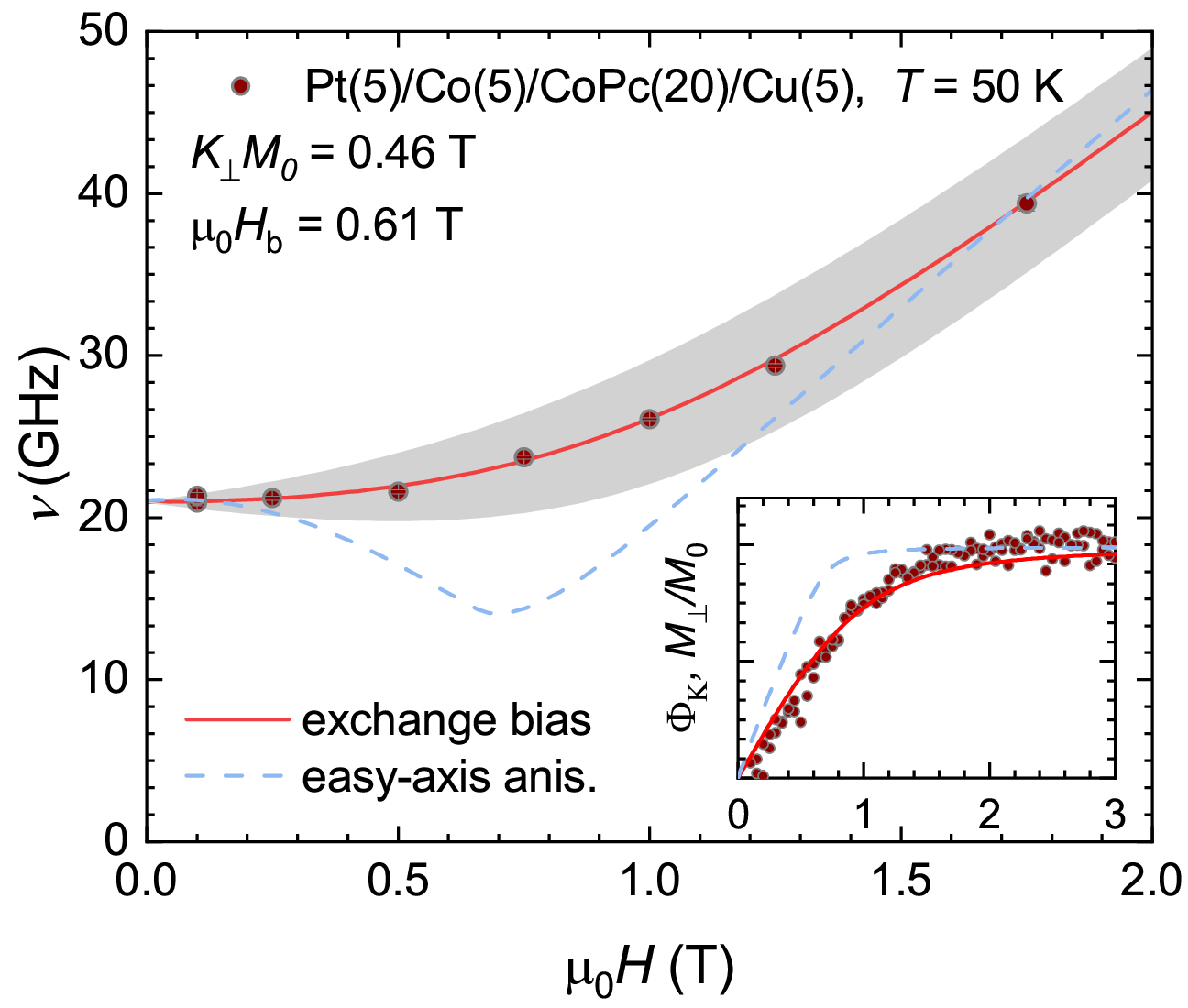}\caption{Comparison of the exchange-bias model \eqref{mod-exb} spin-wave frequency
to the data. The static out-of-plane response is shown in the inset.
The full lines corresponds to the average over a random in-plane exchange-bias
directions distribution, while the gray band to the spread of frequencies
across the distribution. The dashed line correspond to the in-plane
easy-axis model \eqref{mod1a}.\label{fig:fig-bias_model_freq}}
\end{figure}

To model the magnetic field dependence on equal footing through the
full $H$ range we therefore at first neglect the low-$H$ lateral
inhomogeneity. The large low-$H$ frequency suggest the presence of
a rather strong in-plane anisotropy, however, such anisotropy leads
to the spin-wave frequency $H$ dependence that is\textit{ incompatible}
with the observed behavior (see Fig. \ref{fig:fig-bias_model_freq}
and Supplemental, Section \ref{subsec:Uniform-precession-modeling}).
Alternatively, the strong in-plane effective field can be attributed
to an exchange bias \citep{gruber2015exchange,moorsom2020pianisotropy,avedissian2022exchange}
present at the Co/molecules interface. As shown in Fig. \ref{fig:fig-bias_model_freq},
it turns out that an addition of the in-plane exchange-bias-like free-energy
term to the easy-plane-anisotropy model (see Supplemental, Section
\ref{subsec:Uniform-precession-modeling}, Eq. \eqref{mod-exb}) fairly
reproduces\footnote{To take into account zero-field cooling we average over a random in-plane
exchange-bias directions distribution.} both, the dynamic and the out-of-plane static data, with an effective
in-plane exchange bias field of $\mu_{0}H_{\mathrm{b}}\sim0.6$~T
and a reduced (with respect to the high-$T$ behavior) effective easy-plane
anisotropy of $K_{\perp}M_{0}\sim0.5$\,T.

Our data therefore suggest a picture where the presence of the Co/molecules
interface induces rather strong modifications to the interfacial Co
layer(s) in the form akin to an in-plane exchange bias. While the
molecular-induced exchange bias behavior has been observed previously
\citep{gruber2015exchange,boukari2018disentangling,jo2019molecular,moorsom2020pianisotropy},
it might not be intrinsic \cite{avedissian2022exchange} and has not
been observed in some other static experiments \citep{moorsom2014spinpolarized,bairagi2015tuningthe,bairagi2018experimental,benini2022indepth},
in particular in the samples similar to the samples studied in this
work. Moreover, the static in-plane magnetic behavior suggests a correlated
random anisotropy in the absence of any exchange bias term \citep{BeniniShumilin2024}.
The above homogeneous model with a rigid exchange bias therefore appears
inconsistent with the static in-plane behavior.

While the absence of a major static in-plane exchange-bias $H$-shift
can be attributed to a random in-plane distribution of the $H_{\mathrm{b}}$
directions the rigidity of $H_{\mathrm{b}}$ cannot be substantiated.
The effective bulk exchange bias of $\mu_{0}H_{\mathrm{b}}\sim0.6$~T
corresponds to the interface exchange field of the order of $\sim20$
T. To realize such exchange field proximity of a bulk hard (anti)ferromagnet
would be necessary. As the molecules a few layers away from the Co
interface are unlikely to support any magnetic order an interfacial
magnetic state providing the exchange-bias-like free-energy term cannot
be thick and bulk-like.

\begin{figure}
\includegraphics[width=0.9\columnwidth]{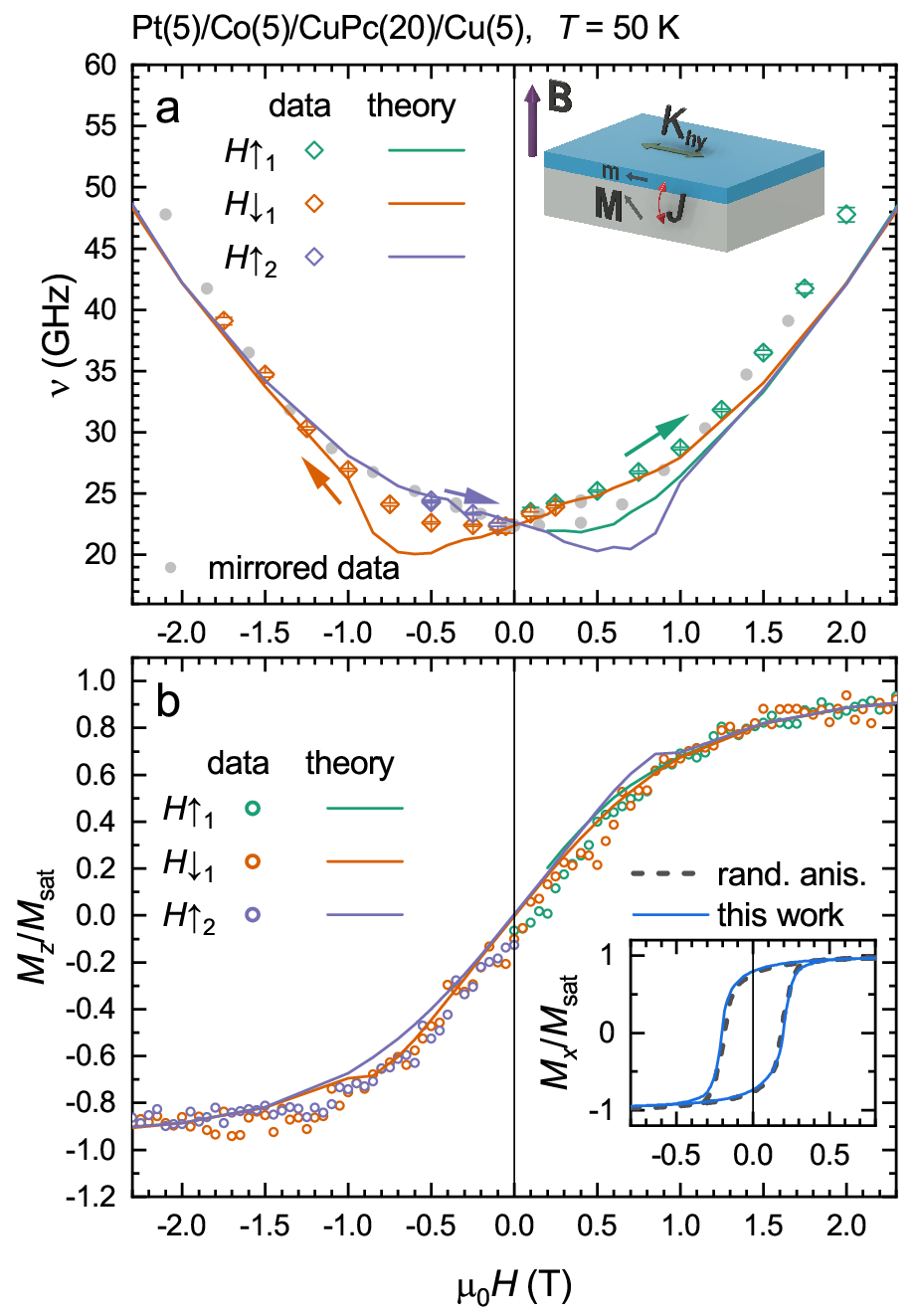}\caption{Comparison of the pseudo-exchange-bias random-anisotropy model (Supplemental,
Section \ref{subsec:Pseudo-exchange-bias-model}) to the experimental
data. \textbf{a }The spin-wave frequency hysteresis after zero-field-cooling
in the CuPc based heterostructure compared to the model. The inset
is a graphical illustration of the model. \textbf{b} The static out-of-plane
response data compared to the model. The inset is a comparison of
the theoretical in-plane hysteresis calculated with the random-anisotropy
model \citep{BeniniShumilin2024} and the pseudo-exchange-bias random-anisotropy
model. The parameters (Supplemental, Section \ref{subsec:Pseudo-exchange-bias-model},
Eq. \eqref{eq:F2D-sim}) for the simulated curves are: $K_{\perp}M_{0}=0.6$~T,
$K_{\mathrm{hy}}m_{0}=14.6$~T, $J/m_{0}=18.3$~T and $dM_{0}/m_{0}=12$,
corresponding to $\alpha=1.2$ and $K_{\mathrm{R}}M_{0}=0.61$~T.
The correlation length is $r_{\mathrm{c}}=5.5\xi_{\mathrm{c}}$.\label{fig:fig-pseudo-bias}}
\end{figure}

There is, however a possibility that the hybridization of the interface
Co layer with molecules leads to formation of a strongly anisotropic
molecular-magnet-like interface magnetism. For example, the high-spin
Co(II) in low-symmetry organic complexes shows unquenched orbital
momentum \citep{craig20153dsingleion,bar2016magnetic} with possibly
large magnetic anisotropy up to several meV per Co ion. A strong covalent
bonding of the surface Co with the molecules combined with charge
transfer might result in a similar magnetic anisotropy. To check whether
such molecular-magnet-like interface magnetism hypothesis can \textit{consistently}
explain the static in-plane and out-of-plane behavior together with
the dynamical response, we extended the previously studied correlated-random-anisotropy
(CRA) model \citep{BeniniShumilin2024} by including a super-paramagnetic
interface magnetic layer (SPIL) that is exchange coupled ($J$) to
the bulk-Co film. The correlated random in-plane anisotropy ($K_{\mathrm{hy}}$)
is assumed to act on the SPIL only (see Supplemental, Section \ref{subsec:Pseudo-exchange-bias-model}
and inset to Fig. \ref{fig:fig-pseudo-bias}a). Further, we assume
that the Néel relaxation time is much faster then the spin-wave frequency
so the local SPIL magnetization always follows the local SPIL effective
field. As shown in Supplemental, Section \ref{subsec:Pseudo-exchange-bias-model}
the SPIL effective field is always large enough to keep the SPIL magnetization
saturated. If the exchange coupling, $J$, is very large the SPIL
magnetization, $\boldsymbol{\mathrm{m}}$, always follows the bulk
Co-film magnetization, $\boldsymbol{\mathrm{M}}$, and the model becomes
equivalent to the original CRA model. On the other hand, when the
effective exchange field in the SPIL, $\mathrm{\boldsymbol{H}_{m,e}}=J\boldsymbol{\mathrm{M}}/\mu_{0}M_{0}m_{0}$,
is comparable to the local anisotropy field, $\mathrm{\boldsymbol{H}_{m,a}}=K_{\mathrm{hy}}\mathrm{\boldsymbol{m}}/\mu_{0}$,
the SPIL magnetization acquires finite angle with respect to $\boldsymbol{\mathrm{M}}$,
 resulting in an effective pseudo-exchange-bias (PEB) free energy
term, which is similar to the exchange-bias term (Supplemental, Section
\ref{subsec:Pseudo-exchange-bias-model}, Eq. \eqref{pb-interpol}).

As shown in Fig. \ref{fig:fig-pseudo-bias} the behavior of such PEB-CRA
model qualitatively matches all, the in-plane and the out-of-plane
static data as well as the magnetic dynamics data. Comparing the theoretical
spin-wave frequency hysteresis (Fig. \ref{fig:fig-pseudo-bias}a)
with the Pt(5)/Co(5)/CuPc(20)/Cu(5) heterostructure data the main
discrepancy is the magnitude of the splitting between the lower and
upper hysteresis branches. This splitting varies also with molecular
species (see Fig. \ref{fig:FC-hyst}). Since, in addition to the model
parameters and $T$, the hysteresis can be affected also by extrinsic
effects such as grain boundaries and the laser excitation during measurement
we believe that the discrepancy is not essential.

The PEB-CRA model is qualitatively consistent also with the field-cooled
dynamical hysteresis data (Fig. \ref{fig:FC-hyst}). However, similar
to the static in-plane hysteresis data in certain Ta-seeded Co/C$_{60}$
based heterostructures \citep{moorsom2020pianisotropy}, strong training
effects are observed during the initial out-of-plane magnetization
saturation and the subsequent hysteresis loops show a magnetic-field
offset of $\sim0.1-0.2$ T. While some training effects are observed
in the simulations the magnetic-field offset is forbidden by the symmetry.
The presence of the magnetic-field offset in the field-cooling experiment
therefore cannot be explained by the PEB-CRA model. One possibility
is that the maximum applied out-of-plane magnetic field in the field-cooled
experiment was too small\footnote{The magnetic field is nearly perpendicular to the heterostructure
plane so the in-plane magnetic-field component in Fig. \eqref{fig:FC-hyst}
did not exceed $\sim0.1$~T, which is comparable to the in-plane
coercitivity \cite{BeniniShumilin2024}.} to fully saturate the in-plane magnetization component in all grains
in the probed volume and the experimental data correspond to a minor
loop.

The data also indicate that the general molecular shape does not strongly
influence the molecules-induced PEB and the characteristic $T$ scale.
This could indicate that the interface is possibly oxidized as proposed
recently \citep{avedissian2022exchange}. In order to check for oxidation
we performed transient MOKE experiments also in a thin film sample
consisting of naturally oxidized 5 nm thick Co. The presence of antiferromagnetic
CoO$_{x}$ at the surface also resulted in low-$T$ spin-wave frequency
hardening. The degree of hardening ($\nu\sim10$~GHz at $\mu_{0}H\sim0$
T) is however significantly smaller with completely different $T$
and $H$ dependencies\footnote{A sharp, $T$-dependent, meta-magnetic transition was observed around
$\mu_{0}H\sim0.8$ T (at $T=50$ K) which will be discussed in detail
elsewhere.} (see Fig. \ref{fig:Frequency-as-function}a, c and Supplemental,
Section \ref{sec:Oxidized-Co-film}, Fig. \ref{fig:CoOx}). The observed
effects in the Co/molecules heterostructures can therefore be reliably
attributed to the molecular chemisorption to the Co-film surface,
although the possible role of oxygen contamination in the chemisorption
remains unclear \citep{avedissian2022exchange}.

The common feature of all the investigated molecules is the presence
of aromatic rings that are predicted to hybridize with Co \citep{chen2010effectof,droghetti2014electronic,bairagi2018experimental}
via their \textgreek{p} orbitals. In the case of Gaq$_{3}$ and C$_{60}$
the molecular geometry is such that only part of the rings are close
enough to the interface to hybridize, while in the case of MPc all
rings can hybridize. The surface density of the hybridized rings is
therefore expected to be different for different molecules. The absence
of large differences of the molecules-induced spin-wave frequency
hardening and the onset $T$ for different molecules suggests that
the effect might be saturated beyond some critical hybridized Co-atoms
surface density, however, further studies are necessary to prove the
hypothesis.

While strong interfacial modification is experimentally hinted also
by the decrease of the saturated magnetic moment, up to $\sim15\%$
upon covering 5 nm thick Co film with 20 nm of C$_{60}$ \citep{moorsom2014spinpolarized,bairagi2018experimental},
the ab-initio calculations \citep{chen2010effectof,droghetti2014electronic,bairagi2018experimental,halder2023theoretical}
generally do not predict such strong effects. Here we should take
into account that our data suggest that the interfacial magnetic state
should involve a correlated state possibly distributed laterally across
several molecules, while the ab-initio calculations are usually limited
to well separated independent molecules and might miss the longer
range correlation effects. A related modification of a few mono-layer
thick Co film anisotropy upon covering with a graphene mono-layer
reported recently \citep{brondintailoring} also suggests that a formation
of correlated C-Co bond arrays might be necessary to achieve the effect.
The characteristic onset $T\sim200$ K, associated with the spin-wave
hardening effect, could therefore be connected with the emergence
of the correlation effects leading to the interface super-paramagnetic
ordering.

\section{Conclusions}

We systematically studied the effect of Co/molecular-semiconductor
interface on the Co-film magnetization dynamics by means of the time-resolved
MOKE spectroscopy in heterostructures based on 5 nm thick Co films
for a range of molecules containing aromatic rings, but having very
different shapes. For all studied molecules we observe a remarkable
hardening of the coherently excited spin-wave frequencies setting-in
below $T\sim200$ K. The spin-wave frequency hardening and the characteristic
$T$-dependencies are not strongly affected by the shape of the molecules.
The data therefore suggest that the presence of aromatic rings that
hybridize with Co at the interface is the common denominator.

The hardening, which is observed also in the absence of the external
magnetic field, implies a rather large in-plane effective field of
$\mu_{0}H_{\mathrm{||eff}}\sim0.9$ T induced by the Co/molecular
interface. The external magnetic field dependence of the static and
dynamic magnetic properties suggests that the effective field is due
to a spatially-correlated random in-plane interfacial pseudo-exchange
bias, as the simpler spatially-correlated random in-plane easy-axis
anisotropy model cannot reproduce the experimental observations. The
interfacial spatially-correlated random pseudo-exchange bias implies
the presence of a rather anisotropic interfacial super-paramagnetism
below $T\sim200$ K that involves the interfacial Co layer, but is
not directly related to the bulk Co ferromagnetism. The emergence
of the anisotropic interfacial super-paramagnetism is therefore tentatively
attributed to a strong chemical modification of the Co interfacial
layer by the chemisorbed molecular layer.

\section{Methods}

\subsection{Time resolved magneto-optical Kerr measurements}

The sample is mounted in a magneto-cryostat with the external magnetic
field $H$ approximately perpendicular to the Co-molecular thin-film
heterostructure as shown in Fig. \ref{fig:Experimental-setup-and}
a. An ultrafast laser pump pulse, ($\tau_{\mathrm{P}}$ < 60 fs, photon-energy
$\hbar\omega=3.1$ eV or 1.55 eV) excites the system inducing a rapid
change of the effective anisotropy field from the thermal equilibrium
value, $H\mathrm{_{a}}$, to a perturbed value, $H\mathrm{_{a}}'$,
where both can be statically affected by the external field, $H$.
As a result the magnetization $M$ starts to coherently precess around
the perturbed equilibrium position while losing the unaligned angular
momentum component by interactions with the system electronic and
lattice excitations. \citep{vankampen2002alloptical} The magnetization
dynamics is probed by using a probe laser pulse (photon-energy $\hbar\omega=1.55$
eV) with a variable time delay with respect to the pump pulse. This
dynamics is related to the one observed in the ferromagnetic resonance
(FMR) experiments and the spin-wave oscillation frequency, \textgreek{n},
is expected to depend on the anisotropy field of the system \citep{vankampen2002alloptical,bigot2005ultrafast}.
This means that this experimental configuration can be used to extract
information about the magnetic anisotropy due to the presence of the
interfacial FM-molecular hybridization.

\subsection{Heterostructures fabrication and characterization}

For the 2-layer heterostructures thin Co films (5~nm) were deposited
by electron beam evaporation on Al$_{2}$O$_{3}$ (0001) substrates
at a room temperature and base pressure of 1.1x10$^{-10}$ mbar with
the deposition rate 0.03 Å/s. The organic layer or the Al layer were
subsequently deposited on top of the Co layer by thermal evaporation
(base pressure 1.1x10$^{-9}$ mbar) at room temperature without breaking
vacuum. The deposition rates were 0.15 Å/s for C$_{60}$, 0.25 Å/s
for Gaq$_{3}$ and 0.1 Å/s for Al.

The 4-layer thin-film structures were grown on 0.5 mm thick (0001)
Al$_{2}$O$_{3}$ substrates. The (111) textured Pt layers of $\sim4$~nm
thickness were grown at 500 °C by means of e-beam evaporation at a
growth rate of $\sim0.1$ Å/s. The substrate was then cooled down
to room temperature and the Co-layer (5~nm) was grown also with e-beam
evaporation technique at a rate of $\sim0.1$ \.{A}/s. Within the
same chamber, organic molecule layers were sublimed onto the Co surface
at a pressure of 5x10$^{-10}$ mbar with a rate of $\sim0.3$ Å/s
and until a thickness of 20~nm was measured through a quartz monitor.
The Cu cap layer (5nm) was magnetron sputtered on top of the thin
film structure. The heterostructures were then structurally characterized
by means of the X-ray reflectivity and Raman spectroscopy.

\section*{Bibliography}

\bibliographystyle{naturemag}
\bibliography{biblio}

\begin{acknowledgments}
The authors acknowledge the financial support of EC projects INTERFAST
(H2020-FET-OPEN-965046). J. S. acknowledges Slovenian Research and
Innovation Agency young researcher funding No. PR-12840.
\end{acknowledgments}

\section*{Author contributions}

T.M., V.A.D. and M.C. developed the hypothesis and coordinated the
experiment. V.A.D. and O.C. defined the heterostructures and coordinated
their fabrication. M.B., R.R., M.R. and O.C. fabricated an characterized
the heterostructures. J.S., H.Z., G.J. and M.B. conducted the TR-MOKE
experiments. J.S. and T.M. analyzed the TR-MOKE data. J.S., A.S.,
V.V.K. and T.M. contributed to modeling. All authors contributed to
the interpretation of the data. T.M. wrote the paper with inputs from
J.A., A.S., V.A.D., O.C. and M.C..

\section*{Competing interests}

Authors declare no competing interests.

\section*{Materials \& Correspondence}

Correspondence and material requests should be addressed to T.M.,
V.A.D. or O.C..

\clearpage{}

\part*{Supplemental}

\global\long\def\thefigure{S\arabic{figure}}%
 
\global\long\def\theequation{S\arabic{equation}}%
 \setcounter{figure}{0} \setcounter{section}{0} \setcounter{equation}{0}

\subsection{Zero-field cooled hysteresis loops\label{sec:Field-cooled-loops}}

\begin{figure}[b]
\includegraphics[width=0.9\columnwidth]{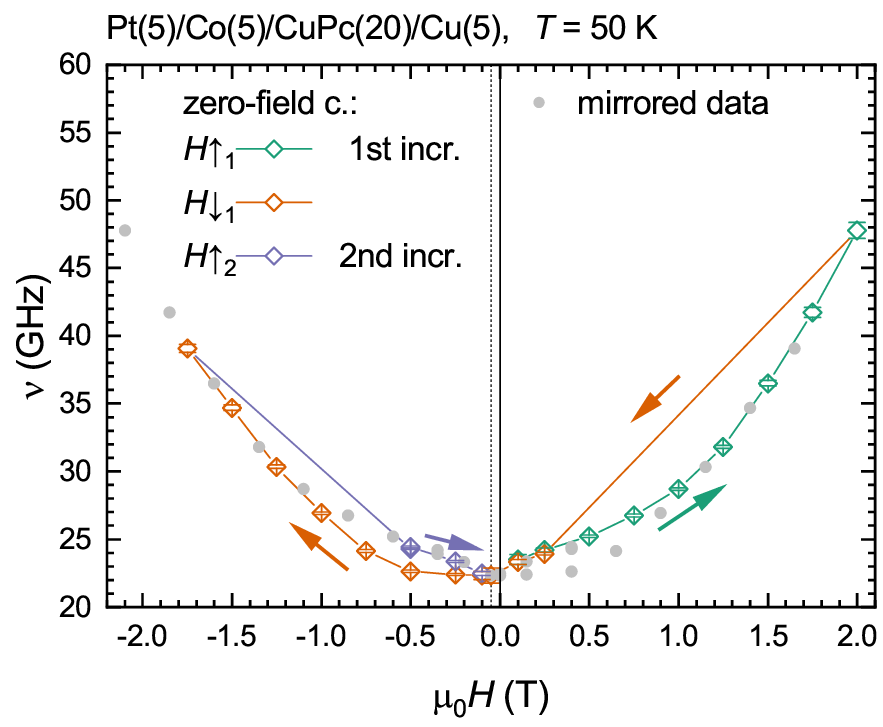}\caption{The spin-wave frequency $H$-dependence in the zero-field cooled case
in the Pt(5)/Co(5)/CuPc(20)/Cu(5) heterostructure. The gray circles
correspond to the data mirrored across the $\mu_{0}H=0.05$~T vertical
line.\label{fig:FCvsZFC}}
\end{figure}

In Fig. \ref{fig:FCvsZFC} we plot the $H$-dependent spin-wave frequency
in the zero-field cooling case in the Pt(5)/Co(5)/CuPc(20)/Cu(5) heterostructure.
A reversible behavior is observed during the initial field sweep to
$\mu_{0}H=2$~T and back to 0\_T, while the subsequent negative field
sweep to $\mu_{0}H=-1.8$~T shows an asymmetric frequency-field dependence
with respect to the positive sweep with an irreversible behavior when
returning towards 0~T. Comparing the original data with the data
mirrored across the $\mu_{0}H=0.05$\,T vertical line reveals a similar
shape as observed in the 2nd loop upon in-plane-field cooling in the
Pt(5)/Co(5)/C$_{60}$(20)/Cu(5) heterostructure (see the main article,
Fig. \ref{fig:FC-hyst}b).

\subsection{Fit function\label{sec:Fit-function}}

\begin{figure}
\includegraphics[width=1\columnwidth]{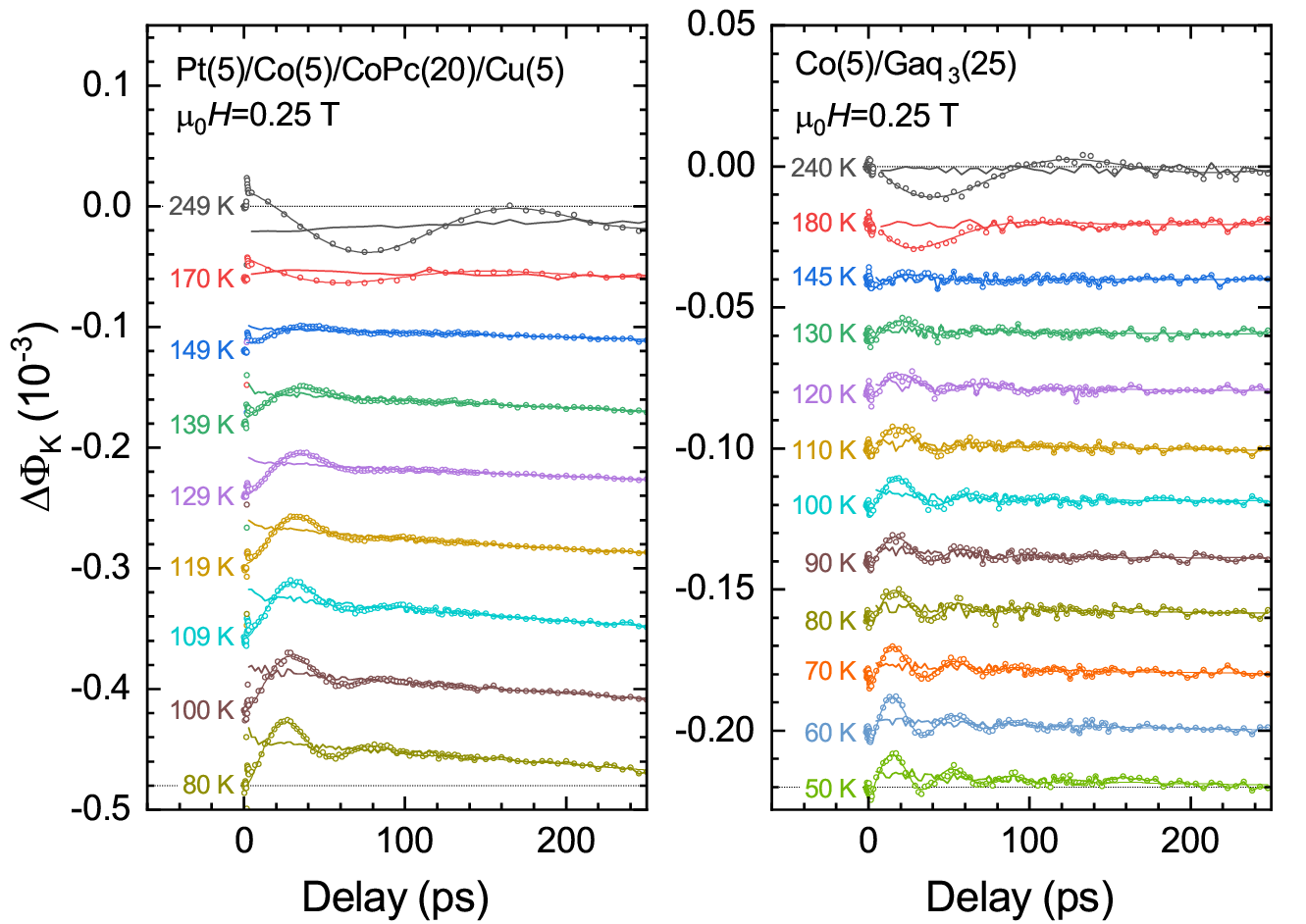}\caption{Examples of transient data with subtracted oscillatory component.
Symbols correspond to data, the thin lines to fits \eqref{eq:fitf}
and the thicker lines to the data with subtracted oscillatory component.
The traces are vertically offset for clarity. The offset is indicated
by the dotted lines for the first and last data sets.\label{fig:Example-of-transient}}
\end{figure}

To extract the oscillation frequency and other parameters we fit the
data ($t>3$ ps) by a damped-cosine function, 
\begin{eqnarray}
f(t) & = & A_{\mathrm{o}}e^{-\gamma t}\cos(2\pi\nu+\delta)\nonumber \\
 &  & +A_{\mathrm{e}1}e^{-\nicefrac{t}{\tau_{1}}}+A_{\mathrm{e}2}e^{-\nicefrac{t}{\tau_{2}}}+y_{0},\label{eq:fitf}
\end{eqnarray}
where $A_{\mathrm{o}}$, $\nu$, $\gamma$ and $\delta$ correspond
to the amplitude, oscillation frequency, damping and phase shift,
respectively. The last three terms represent the non-coherent background,
where in most cases the second exponential term could be set to zero
and $\tau_{1}$ fixed to a nanosecond, producing a virtually linear
$t$-dependent background within the measurement delay range.

\subsection{Fluence dependence}

\begin{figure}
\includegraphics[width=0.8\columnwidth]{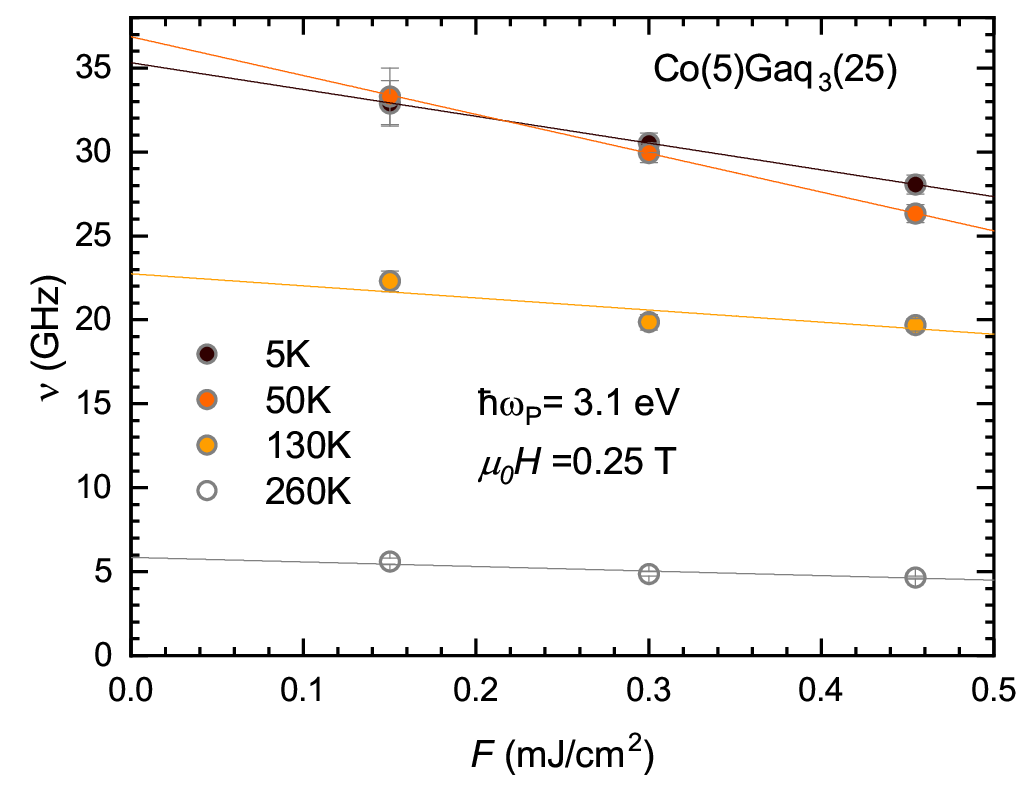}\caption{Pump fluence dependence of the frequency in a Co(5)/Gaq$_{3}$(20)
heterostructure at different $T$. The lines correspond to linear
fits. \label{fig:F-dep-Gaq3}}
\end{figure}

The fluence dependence was studied systematically in a Co(5)/Gaq$_{3}$(20)
heterostructure only. The frequency shows softening with increasing
$F$ as shown in Fig. \ref{fig:F-dep-Gaq3}. The $F$ dependence in
other heterostructures is qualitatively similar, but was not systematically
studied in the present work.

\subsection{Oxidized Co film\label{sec:Oxidized-Co-film}}

\begin{figure}
\includegraphics[width=0.9\columnwidth]{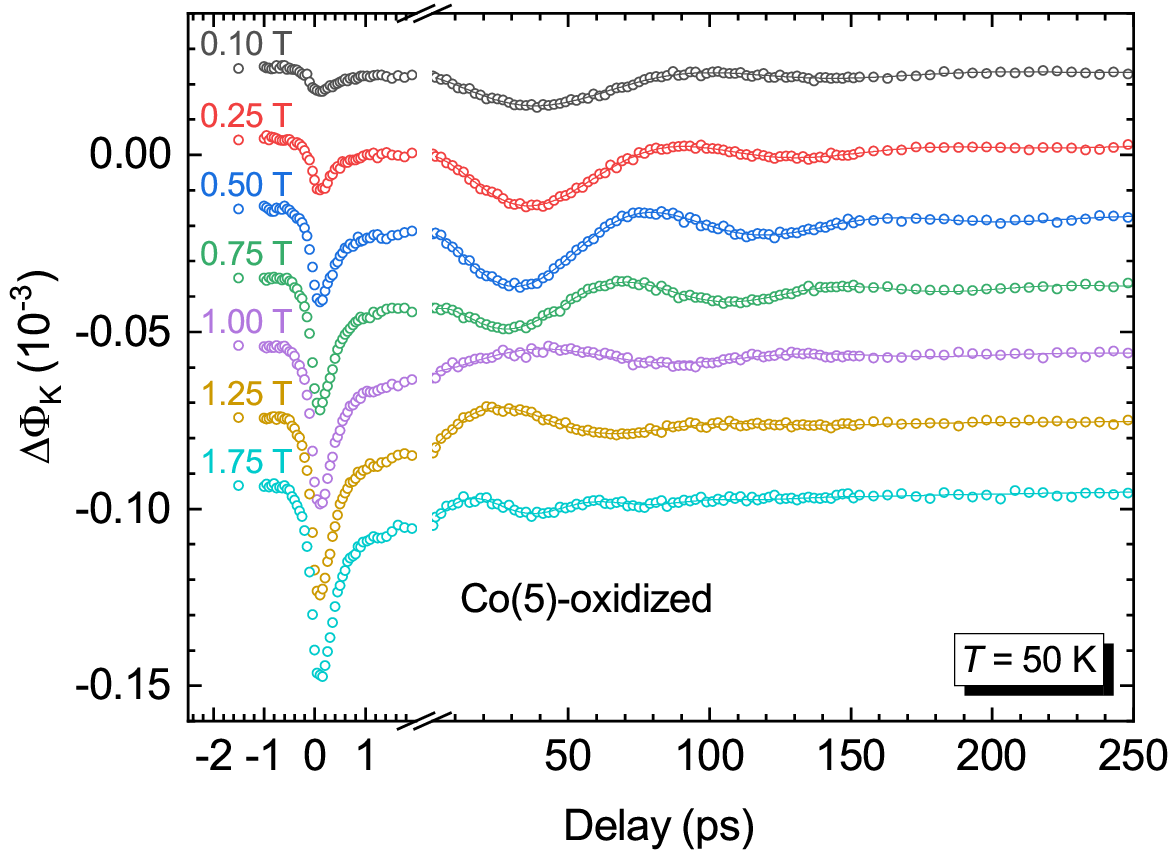}\caption{Transient magneto-optical Kerr rotation in naturally oxidized Co(5)
film. Note the change of spin-wave phase (sign) between $\mu_{0}H=0.75$
T and $1$ T. The thin lines are fits \eqref{eq:fitf}. The traces
are vertically shifted for clarity.\label{fig:CoOx}}
\end{figure}

In Fig. \ref{fig:CoOx} we show the low-$T$ $H$-dependent transient
magneto-optical Kerr rotation in naturally oxidized Co(5) film. A
meta-magnetic transition is observed between $\mu_{0}H=0.75$ T and
$1$ T at $T=50$~K, where the coherent spin-wave response shows
change of phase. The frequency as a function of $H$ (shown in main
text, Fig. \ref{fig:Frequency-as-function}a) shows a frequency jump
leading to a non-monotonous behavior across the meta-magnetic transition.

\subsection{Transient reflectivity\label{subsec:Transient-reflectivity}}

\begin{figure}
\includegraphics[width=0.8\columnwidth]{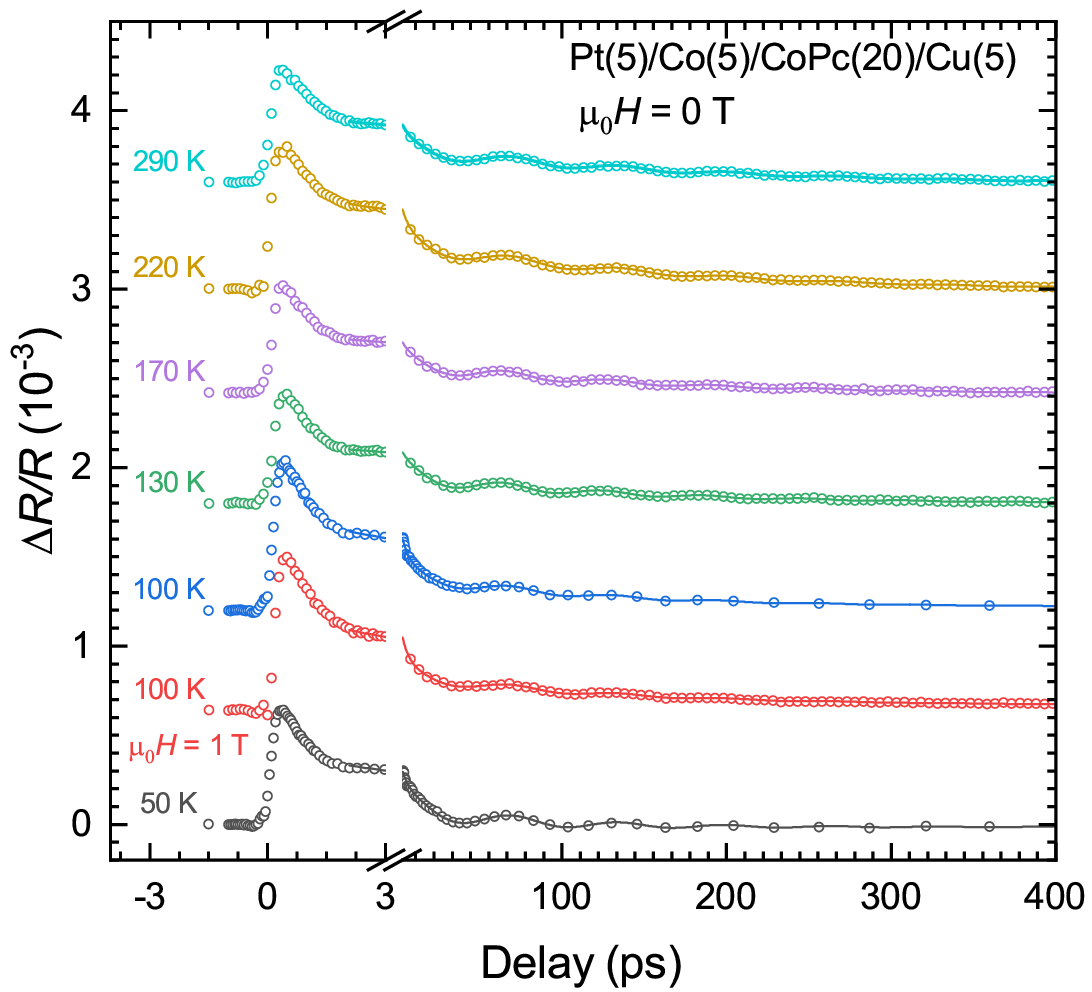}\caption{Transient reflectivity in Pt(5)/Co(5)/CoPc(20)/Cu(5) heterostructure
as a function of $T$ and $H.$ The lines correspond to the fit \eqref{eq:fitf}.\label{fig:Transient-reflectivity}}
\end{figure}

The transient reflectivity does not show any significant dependence
on $T$ and $H$ as show in Fig. \ref{fig:Transient-reflectivity}
for the case of Pt(5)/Co(5)/CoPc(20)/Cu(5) heterostructure. The coherent
oscillations, which do not depend significantly on $T$ and $H$ are
assigned, to the Brillouin scattering on a coherently-excited acoustic
phonon waves \citep{thomsen1986picosecond} propagating in the substrate.
They can be detected in some heterostructures because they are not
completely opaque.

\subsection{Uniform precession modeling\label{subsec:Uniform-precession-modeling}}

\begin{figure}
\includegraphics[width=0.8\columnwidth]{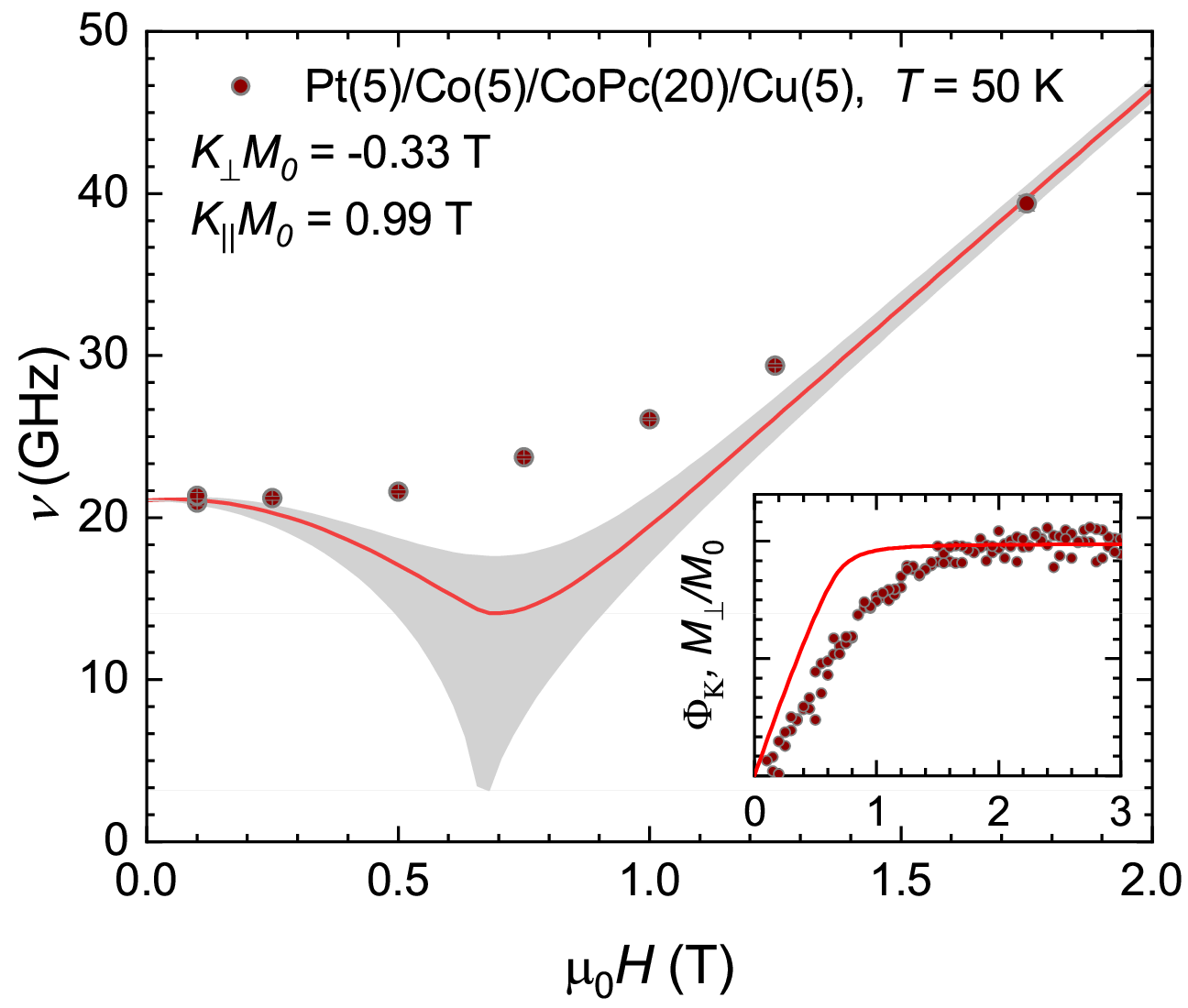}\caption{Comparison of the model \eqref{mod1a} spin-wave frequency to the
data. The static out-of-plane response is shown in the inset. The
parameters were chosen to simultaneously fit the low-$H$ and large-$H$
frequency. The full lines corresponds to the uniform average over
full ${\bf e}_{||}$ polar angle, while the gray band to the spread
of frequencies across the ${\bf e}_{||}$ polar angle. \label{fig:in-plane-anis}}
\end{figure}

Both static magnetization and the oscillation frequencies in the reference
sample can be described by the simple easy-plane-anisotropy model,
where the magnetization is assumed to be homogeneous with the free
energy density, 
\begin{equation}
F=-\mu_{0}{\bf H}\cdot{\bf M}+\frac{K_{\perp}}{2}({\bf M}\cdot{\bf e}_{\perp})^{2},\label{mod1}
\end{equation}
where ${\bf M}$ corresponds to the magnetization, ${\bf e}_{\perp}$
to the unit vector normal to the plane, $K_{\perp}$ the easy-plane/hard-axis
anisotropy constant and ${\bf H}$ the external magnetic field, which
is tilted by small angle $\Theta$ from the film normal. The model
shows quantitative agreement with the Co(5)/Al(3) data (see Fig.~\ref{fig:Frequency-as-function}
a) with the parameters $\Theta=6.2^{\circ}$, $K_{\perp}M_{0}=1.08$\,T
and the gyromagnetic ratio $\gamma=27$\,GHz/T.

A simple extension to the free energy density \eqref{mod1} is the
addition of an in-plane easy axis term,

\begin{equation}
F=-\mu_{0}{\bf H}\cdot{\bf M+}\frac{K_{\perp}}{2}({\bf M}\cdot{\bf e}_{\perp})^{2}-\frac{K_{||}}{2}({\bf M}\cdot{\bf e}_{||})^{2},\label{mod1a}
\end{equation}
where ${\bf e}_{||}$ corresponds to an in-plane unit vector and $K_{||}$
to the easy-axis anisotropy constant. Due to the external magnetic
field tilt \eqref{mod1a} the corresponding spin-wave frequency depends
on the ${\bf e}_{||}$ polar angle. The behavior of such model is
incompatible with the experimental data, irrespective of the ${\bf e}_{||}$
polar angle, as shown in Fig. \ref{fig:in-plane-anis}.

Alternatively, the in-plane anisotropy term can be replaced by an
exchange bias term,

\begin{equation}
F=-\mu_{0}{\bf H}\cdot{\bf M+}\frac{K_{\perp}}{2}({\bf M}\cdot{\bf e}_{\perp})^{2}-\mu_{0}H_{b}{\bf e}_{||}\cdot{\bf M},\label{mod-exb}
\end{equation}
where $H_{b}$ corresponds to the volume-averaged interfacial exchange
bias field magnitude. As shown in Fig. \ref{fig:fig-bias_model_freq}
such model, despite the simplicity, rather well fits the data. Moreover,
the behavior is not strongly sensitive to the exchange-bias field
polar angle.

\subsection{Pseudo-exchange-bias random-anisotropy model\label{subsec:Pseudo-exchange-bias-model}}

The present experiments show that the magnetic field dependence of
the spin-wave frequency is incompatible with the quadratic in-plane
anisotropy term in the free energy \eqref{mod1a}. The exchange bias
term in model \eqref{mod-exb}, which is compatible with the data,
is typically produced by bulk-like layers of antiferromagnetic materials
with sublattice magnetizations that follow neither the external magnetic
field nor the biased ferromagnetic layer magnetization. However, when
the thickness of the antiferromagnetic layer is decreased its magnetic
state is expected to soften, starting to depend on the magnetization
of the ferromagnetic layer and, as a consequence, the external magnetic
field. As such it cannot provide a rigid exchange-bias field assumed
in model \eqref{mod-exb}.

In the present case it is not plausible to presume that the bulk of
the organic layer in our samples is magnetically ordered. Any magnetic
order, induced by the Co-molecules hybridization has to be spatially
confined in a thin layer at the Co-molecules interface. Moreover,
the exchange bias hypothesis is inconsistent with the in-plane static
hysteresis measurements and the corresponding theory \cite{BeniniShumilin2024}.

To resolve this apparent inconsistency between the dynamical response,
the in-plane hysteresis behavior and the device composition we introduce
a pseudo-exchange-bias random-anisotropy model. We consider it as
a minimal model that allows to consistently explain experimental findings
taking into account the samples structure. The model is based on the
following assumptions:
\begin{itemize}
\item The interfacial Co layer is strongly hybridized with molecules leading
to a significant modification of its magnetic properties. While having
an exchange interaction with the bulk Co, this layer is otherwise
super paramagnetic. It means that it can be described by a classical
2D magnetic-moment density ${\bf m}$ but has no coercitivity nor
exchange stiffness of its own. The direction of ${\bf m}({\bf r})$
at any point on the interface ${\bf r}$ is controlled only by its
local anisotropy and the magnetization of the bulk of the Cobalt at
the same point ${\bf r}$.
\item The surface density of the hybridized-Co-layer magnetic moment is
much smaller than that of the bulk Co and is not detected in experiment.
\item The hybridized-Co layer has a rather strong random anisotropy characterized
by the spatial correlation length $r_{\mathrm{c}}\sim10\,{\rm nm}$,
similar to the theory in Ref. \cite{BeniniShumilin2024}.
\item The exchange interaction of the hybridized-Co layer with the bulk
Co is reduced and is comparable to the anisotropy.
\end{itemize}
The model is illustrated in the main text, Fig. \ref{fig:fig-pseudo-bias}a.

\begin{figure}[th]
\centering \includegraphics[width=0.8\columnwidth]{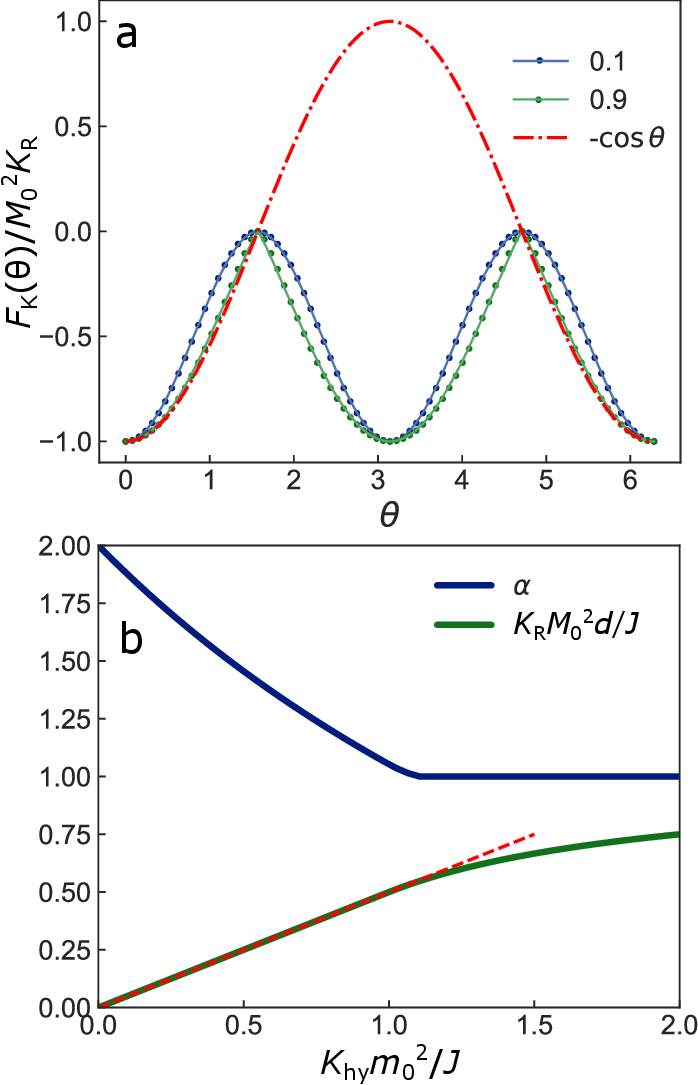} \caption{\textbf{a} The effective anisotropy free-energy term, $F_{\mathrm{K}}({\bf M})$,
as a function of the angle $\theta$ between ${\bf M}$ and ${\bf \boldsymbol{e}_{R}}$
calculated numerically (dots) and with Eq.~(\ref{pb-interpol}) (solid
lines) for the values of $K_{\mathrm{hy}}m_{0}^{2}/J$ shown in the
legend. The red dot-dashed line corresponds to $-\cos\theta$ that
describes the exchange bias term in Eq. \eqref{mod-exb}. \textbf{b}
The values of $K_{R}$ and $\alpha$ obtained by the approximation
for different relations between $K_{\mathrm{hy}}$ and $J$. The red
dashed line shows the asymptotic $K_{\mathrm{R}}=K_{\mathrm{hy}}m_{0}^{2}/2dM_{0}^{2}$.}
\label{fig:pb}
\end{figure}

The exchange-energy surface density between the bulk-Co layer and
the the thin hybridized-Co layer equals $-J{\bf M\cdot}{\bf m}/M_{0}m_{0}$
where $m_{0}=|{\bf m}|$ is the magnitude of the hybridized-layer
magnetic-moment \textit{surface density}, ${\bf m}$, and $M_{0}=|{\bf M}|$
is the absolute value of the bulk-Co layer magnetization, ${\bf M}$.
We assume that the bulk-Co layer is thin compared to the Co exchange
stiffness which allows us to neglect the variation of ${\bf M}$ normal
to the layer. The total free-energy surface density can then be written
as: 
\begin{eqnarray}
F_{\mathrm{2D}} & = & d\left[\mu_{0}\xi_{\mathrm{c}}^{2}\sum_{\beta}\frac{(\nabla M_{\beta})^{2}}{2}-\mu_{0}{\bf H}\cdot{\bf M}+\frac{K_{\perp}}{2}({\bf M}\cdot{\bf e}_{\perp})^{2}\right]\nonumber \\
 &  & -\frac{K_{\mathrm{hy}}}{2}({\bf m}\cdot{\bf \boldsymbol{e}_{R}})^{2}-\mu_{0}{\bf H}\cdot{\bf m}-\frac{J}{M_{0}m_{0}}{\bf M\cdot}{\bf m}.\label{pb-main}
\end{eqnarray}
Here $\xi_{\mathrm{c}}$ is the bulk-Co-layer exchange length while
$\beta$ enumerates the in-plane, $\left\{ x,y\right\} $, Cartesian
coordinates and $d$ is the bulk-Co-layer thickness. $K_{\perp}$
and $K_{\mathrm{hy}}$ are the bulk-Co-layer hard-axis anisotropy
and the hybridized-Co-layer random anisotropy constants, respectively,
while ${\bf \boldsymbol{e}_{R}}$ is a random in-plane unit-vector
field with the correlation length $r_{\mathrm{c}}$. $d$ is the thickness
of the bulk-Co layer.

Eq.~(\ref{pb-main}) shows that the hybridized-Co-layer action on
the bulk-Co layer can be described with an effective exchange field,
\begin{eqnarray}
{\bf H}_{M,e} & = & \frac{1}{\mu_{0}d}\left[\frac{\partial\left(F_{\mathrm{2D}}\right)}{\partial{\bf M}}-\left.\frac{\partial\left(F_{\mathrm{2D}}\right)}{\partial{\bf M}}\right|_{J=0}\right]\\
 & = & \frac{J}{\mu_{0}M_{0}m_{0}}\frac{{\bf m}}{d}.
\end{eqnarray}
As it will be shown later, the typical value for this field is ${\bf H}_{\mathrm{M,e}}\sim1$~T
in our case. The action of the bulk-Co layer on the hybridized-Co
layer can be described by the effective exchange field, 
\begin{equation}
{\bf H}_{\mathrm{m,e}}=\frac{J}{\mu_{0}M_{0}m_{0}}{\bf M}.
\end{equation}
This field is much larger than ${\bf H}_{\mathrm{M,e}}$: $|{\bf H}_{m,e}|/|{\bf H}_{M,e}|=dM_{0}/m_{0}$.
In our samples the bulk of the Co consists of approximately 12 layers
making $|{\bf H}_{\mathrm{m,e}}|\sim12\,$T. As a result, the dynamics
of the hybridized-Co layer is much faster than the dynamics of the
bulk-Co layer and when studying the later we can safely assume that
${\bf m}$ always adiabatically adapts to ${\bf M}$ according to
Eq. \eqref{pb-main}. The large magnitude of ${\bf H}_{\mathrm{m,e}}$
also allows us to neglect the external field term ${\bf \mu_{0}{\bf H}\cdot{\bf m}}$
in Eq.~(\ref{pb-main}). Note that the relation $|{\bf H}_{\mathrm{m,e}}|\gg|{\bf H}_{\mathrm{M}}|$
makes our model different from Refs.~\cite{wigen1962dynamic,lubitz1998temperature}
where the relation was opposite.

The fast dynamics of the hybridized-Co layer magnetization allows
us to exclude ${\bf m}$ from Eq.~(\ref{pb-main}) and reduce the
problem to the dynamics of ${\bf M}$ with an effective anisotropy
term $F_{\mathrm{K}}({\bf M})$ in the expression for the free energy
density of the Co layer, which can be introduced into Eq.~(\ref{pb-main}),
by replacing the terms explicitly containing ${\bf m}$. The expression
for $F_{K}({\bf M})$ should be derived by minimizing Eq.~(\ref{pb-main})
with respect to ${\bf m}$ at any given ${\bf M}$ and substituting
${\bf m}$ in Eq.~(\ref{pb-main}) with the resulting ${\bf m}_{\mathrm{min}}({\bf M})$.
The expression for $F_{\mathrm{K}}$,
\begin{equation}
F_{\mathrm{K}}({\bf M})=-\frac{K_{\mathrm{hy}}}{2d}({\bf m}_{\mathrm{min}}({\bf M})\cdot{\bf \boldsymbol{e}_{R}})^{2}-\frac{J}{dM_{0}m_{0}}{\bf M\cdot}{\bf m}_{\mathrm{min}}({\bf M}),
\end{equation}
is generally more complex than the simple quadratic term, $K_{\mathrm{||}}({\bf M}\cdot{\bf \boldsymbol{e}_{||}})^{2}/2$
used in Eq. \eqref{mod1a}. We were not able to rigorously derive
it analytically, however it can be easily calculated numerically.
We also found that in all the cases it can be very well approximated
by a phenomenological equation: 
\begin{equation}
F_{\mathrm{K}}({\bf M})\simeq-M_{0}^{2}K_{\mathrm{R}}\left|\frac{{\bf M}\cdot{\bf \boldsymbol{e}_{R}}}{M_{0}}\right|^{\alpha}.\label{pb-interpol}
\end{equation}
Here $K_{R}$ describes the strength of the effective random anisotropy
applied to the bulk Co and $1<\alpha<2$ is some power. This approximation
is shown in Fig.~\ref{fig:pb}a where the points describe the numeric
calculation of $F_{K}$ and the solid lines correspond to Eq.~(\ref{pb-interpol}).
Fig.~\ref{fig:pb}b also shows the values of $K_{\mathrm{R}}$ and
$\alpha$ obtained by the approximation for different relations between
$J$ and $K_{\mathrm{hy}}$.

There are two clear limiting cases for $F_{K}$. When the exchange
interaction between bulk of the Co and the hybridized-Co layer is
strong $J\gg K_{\mathrm{hy}}m_{0}^{2}$, $\alpha=2$ and $K_{R}=K_{\mathrm{hy}}m_{0}^{2}/2dM_{0}^{2}$.
It means that the surface anisotropy is equally re-distributed across
the bulk-Co layer by the strong exchange interaction. This situation
reproduces the common quadratic anisotropy term. In the opposite case,
$K_{\mathrm{hy}}m_{0}^{2}\gg J$, $\alpha$ becomes equal to 1 and
$K_{R}=J/dM_{0}^{2}$. In this case $F_{K}\propto-|{\bf M}\cdot{\bf \boldsymbol{e}_{\boldsymbol{R}}}|$
which can be called a pseudo-bias anisotropy (the standard exchange
bias would correspond to $F_{K}\propto-{\bf M}\cdot{\bf \boldsymbol{e}_{R}}$
without the "absolute value"). Physically it means that the anisotropy
$K_{\mathrm{hy}}$ always keeps ${\bf m}$ directed parallel to ${\bf \boldsymbol{e}_{R}}$,
therefore it acts on the bulk-Co layer similarly to the exchange bias.
However, it can flip between $+{\bf \boldsymbol{e}_{R}}$ and $-{\bf \boldsymbol{e}_{R}}$
directions, therefore, phenomenologicaly Eq.~(\ref{pb-interpol})
with, $\alpha=1$, does not break the time-reversal symmetry and describes
an anisotropy, not a bias.

It is interesting that while the limit $K_{R}=J/dM_{0}^{2}$ is reached
only for very high values of $K_{\mathrm{hy}}$, the pseudo-exchange
bias behavior ($\alpha\approx1$) starts from $K_{\mathrm{hy}}m_{0}^{2}\sim J$.
In Fig.~\ref{fig:pb}a we compare $F_{K}/K_{\mathrm{R}}M_{0}^{2}$
calculated for $K_{\mathrm{hy}}m_{0}^{2}/J=0.9$ with $-\cos(\theta)$
that would correspond to the standard exchange bias. Already at this
value of $K_{\mathrm{hy}}$ the exchange bias behavior is reproduced
for all values of $\theta$ where $\cos\theta>0$.

To simulate the curves in Fig. \ref{fig:fig-pseudo-bias} we therefore
take the free energy density,
\begin{eqnarray}
F_{\mathrm{}} & = & \mu_{0}\xi_{\mathrm{c}}^{2}\sum_{\beta}\frac{(\nabla M_{\alpha})^{2}}{2}-\mu_{0}{\bf H}\cdot{\bf M}+\frac{K_{\perp}}{2}({\bf M}\cdot{\bf e}_{\perp})^{2}\nonumber \\
 &  & -M_{0}^{2}K_{\mathrm{R}}\left|\frac{{\bf M}\cdot{\bf \boldsymbol{e}_{R}}}{M_{0}}\right|^{\alpha},\label{eq:F2D-sim}
\end{eqnarray}
map it to a 2D lattice and first calculate the static ${\bf M}{}_{\mathrm{stat}}(\boldsymbol{\mathrm{r}},{\bf H})$
following the procedures described in Ref. \cite{BeniniShumilin2024}
for a number of different spatially correlated random fields ${\bf \boldsymbol{e}_{R}}(\boldsymbol{\mathrm{r}})$.
The dynamical response is then simulated by calculating the Landau-Lifshitz-Gilbert
time evolution starting with ${\bf M}{}_{\mathrm{stat}}(\boldsymbol{\mathrm{r}},{\bf H})$
using Eq. \ref{eq:F2D-sim} with slightly perturbed parameter, $K_{\mathrm{R}}\rightarrow K_{\mathrm{R}}+\Delta K_{\mathrm{R}}$
at each ${\bf H}$ and $\boldsymbol{{\bf \boldsymbol{e}_{R}}}(\boldsymbol{\mathrm{r}})$.

\end{document}